\documentclass[%
reprint,
amsmath,amssymb,
aps,
]{revtex4-2}

\usepackage{lipsum}
\usepackage{bm}
\usepackage{amssymb}
\usepackage{amstext}
\usepackage{amsmath}
\usepackage{mathrsfs}
\usepackage{graphicx}
\usepackage{color}
\usepackage{amsthm}
\usepackage{floatrow}
\usepackage{array}

\usepackage{listings}
\usepackage{booktabs}
\usepackage{hyperref}

\begin{document}
	\preprint{APS/123-QED}
	\title{Fluid-network relations: decay laws meet with spatial self-similarity, scale-invariance, and control scaling}
	\thanks{Correspondence should be addressed to Pei Sun and Yizhou Xu.}%
	
	\author{Yang Tian}
\email{tyanyang04@gmail.com}
\affiliation{Infplane AI Technologies Ltd, Beijing, 100080, China}
\affiliation{Laboratory of Computational Biology and Complex Systems, City University of Macau, Macau, 999078, China}
\affiliation{Faculty of Health and Wellness, City University of Macau, Macau, 999078, China}
\affiliation{Faculty of Data Science, City University of Macau, Macau, 999078, China}
 
\author{Pei Sun}%
\email{peisun@cityu.edu.mo}
\affiliation{Laboratory of Computational Biology and Complex Systems, City University of Macau, Macau, 999078, China}
\affiliation{Faculty of Health and Wellness, City University of Macau, Macau, 999078, China}

\author{Yizhou Xu}%
\email{yizhou.xu.cs@gmail.com}
\affiliation{Infplane AI Technologies Ltd, Beijing, 100080, China}
\affiliation{\'Ecole Polytechnique F\'ed\'erale de Lausanne, Lausanne, 1015, Switzerland}

	
	
	
\begin{abstract}
Diverse implicit structures of fluids are discovered lately, providing opportunities to study the physics of fluids applying network analysis. Although considerable works devote to identifying informative network structures of fluids, we are limited to a primary stage of understanding what
kinds of information can these identified networks convey about fluids. An essential question is how the mechanical properties of fluids are embodied in the topological properties of networks or vice versa. Here, we tackle this question by revealing a set of fluid-network relations that quantify the interactions between fundamental fluid properties (e.g., kinetic energy and enstrophy decay laws) and defining network characteristics (e.g., spatial self-similarity, scale-invariance, and control scaling). We first analyze spatial self-similarity in its classic and generalized definitions, which respectively reflect whether vortical interactions or their spatial imbalance extents are self-similar in fluid flows. The deviation extents of networks from self-similar states exhibit power-law scaling behaviours with respect to fluid properties, suggesting the diversity among vortex as an indispensable basis of self-similar fluid flows. Then, the same paradigm is adopted to investigate scale-invariance using renormalization groups, which reveals that the breaking extents of scale-invariance in networks, similar to those of spatial self-similarity, follow power-law scaling with respect to fluid properties. Finally, we define a control problem on networks to study the propagation of perturbations through vortical interactions over different ranges. The minimum cost of controlling vortical networks exponentially scales with range diameters (i.e., control distances), whose growth rates experiences temporal decays. We show that this temporal decay speed is fully determined by fluid properties in power-law scaling behaviours. In sum, all these discovered fluid-network relations sketch a picture where we can study the implicit structures of fluids and quantify their interactions with fluid dynamics.

\end{abstract}
\maketitle

\section{Introduction}\label{Sec1}

Most of our knowledge about fluids over decades can be summarized as the understanding of their dynamics \cite{kundu2015fluid} rather than structures (e.g., consider those in solids). This is because fluids do not inherently form explicit structures. It is only in the last few years that scientists have started to comprehend the implicit structures of fluids (i.e., specific observable relations among fluid elements that constrain dynamics) \cite{nair2015network,taira2016network}. These latent structures, characterized by complex networks, provide opportunities to study the physics of vortical flows and turbulence using network analysis \cite{iacobello2021review}.

Although complex networks naturally occur in different fields \cite{costa2011analyzing}, identifying informative network structures for fluids is crucial and non-trivial. To date, extensive mappings from fluids to networks have been explored from the Eulerian \cite{nair2015network,taira2016network,krishnan2019emergence,kawano2023complex,scarsoglio2016complex,iacobello2018spatial,tandon2023multilayer,donges2009complex,zhou2015teleconnection,tupikina2016correlation,krueger2019quantitative}, the Lagrangian \cite{nair2015network,taira2016network,meena2018network,scarsoglio2016complex,iacobello2018spatial,tandon2023multilayer,schlueter2019model,iacobello2019lagrangian,perrone2020wall,perrone2021network}, and the time series \cite{murugesan2015combustion,guan2023multifractality,kaiser2014cluster,iacobello2021review} perspectives, revealing diverse possibilities to analyze the dynamic couplings among fluid elements. In Appendix \ref{ASec1}, we present a comprehensive summary of these network identification techniques. Among these methods, we choose to extract the network structures of vortical interactions, which serves as a mainstream approach in the Eulerian view and enables us to trace the chaotic motions of vortex to underlying networks \cite{nair2015network,taira2016network}.

In this work, we take a step further to explore what kinds of information can these extracted networks convey about the nature of fluids. In pioneer studies, this question has only been tackled in a primary stage. For instance, the vortical networks of isotropy turbulence in an Eulerian view are shown as scale-free, implying that isotropy turbulence lacks a characteristic vortical scale and features a self-similar spatial distribution of vortex \cite{taira2016network}. The decay of turbulence kinetic energy across time, i.e., the vanishing of small vortex, can be reflected by the breaking of spatial self-similarity. From a network science perspective, the above finding suggests that the vortical distribution of isotropy turbulence is robust against random perturbations and such robustness fades away with kinetic energy decay \cite{taira2016network}. Isotropy turbulence with strong robustness cannot be significantly disturbed unless targeted perturbations accurately act on the hubs \cite{cimini2019statistical} of vortical networks \cite{taira2016network}, which enlightens flow control engineering and attracts a series of works on fast network extraction \cite{bai2019randomized} and influential hub identification \cite{meena2021identifying,yeh2021network}. Other inspiring findings can be seen in jets \cite{manshour2015fully,charakopoulos2014application,kobayashi2019spatiotemporal}, plumes \cite{takagi2017nonlinear,takagi2018dynamic,tokami2020spatiotemporal,takagi2018effect}, wakes \cite{tao2019modified,wu2020evolution,nair2018networked}, vortical flows \cite{meena2018network,schlueter2017coherent}, and wall turbulence \cite{iacobello2018spatial,iacobello2019lagrangian}. The main challenge faced by these pioneer works, as discussed in Refs. \cite{iacobello2021review,taira2016network}, is the elusive relation between fluid properties and their network counterparts.

In response to this challenge, we aim at discovering the statistical physics correspondence relations between fluids and their network representations, which are referred to as fluid networks for convenience. Our works are summarized as the following:
\begin{itemize}
    \item[(1) ]As a continuation of the work in Ref. \cite{taira2016network}, we first analyze the statistical correspondence between the decay laws (e.g., kinetic energy and enstrophy decays) of turbulent flows and the breaking of spatial self-similarity in fluid networks in Sec. \ref{Sec3}. 
     \item[(2) ] Apart from the classic definition of spatial self-similarity rooted in degree distributions, we extend our analysis into the generalized self-similarity established on the degree–degree distance \cite{meng2023scale,zhou2020power} in Sec. \ref{Sec4}. The distance distribution enables us to study how the structure imbalance of fluid networks exhibits spatial self-similarity and how this self-similarity is regulated by fluid decay laws.
    \item[(3) ] Beyond the self-similar property, we further explore how fluid decay laws regulate the scale-invariance property of fluid networks in Sec. \ref{Sec5}. Note that scale-invariance is not equivalent to spatial self-similarity for complex networks. The former refers to the strict symmetry under scale transformations \cite{villegas2023laplacian,tian2024fast} while the latter corresponds to the similarity between a local region and its sub-regions on a given scale (see Appendix \ref{ASec2} for details). By investigating scale-invariance, we can relate the fluctuations of turbulent flows across multiple scales.
    \item[(4) ] Parallel to the analysis of scale-invariance, we also explore the correspondence relations between other kinds of network scaling and the decay laws of turbulence properties in Sec. \ref{Sec6}. Specifically, we show how these decay laws are reflected by the scaling of network control costs (i.e., the energy cost of controlling a certain set of units) \cite{klickstein2017energy,klickstein2018control}.
\end{itemize}

We uniformly refer to the statistical connections discovered in these four aspects as fluid-network relations. These relations quantify how the decay laws of fluids (e.g., kinetic energy and enstrophy decays) are reflected by the statistical physics characteristics of fluid networks or vice versa. Although we primarily implement our analysis on decaying and forced turbulent flows under different viscosity and velocity conditions, the same idea can be readily generalized to other kinds of fluids to study these fluid-network relations.

\section{Analysis framework}\label{Sec2}

In our work, we generate freely decaying and forced turbulent flows (i.e., Kolmogorov flows) under different conditions of viscosity, $\nu$, and velocity bound, $u_{m}$. All the computational details of fluid generation are presented in Appendix \ref{ASec3} to ensure the reproducibility. There are $18$ combinations of $\left(\nu,u_{m}\right)$ implemented in our experiments, each of which corresponds to a realization of decaying turbulence and a realization of forced turbulence during a time interval of $\left[0,125\right]$. After sampling $50$ time frames from each realization, we utilize the pipeline introduced in Appendix \ref{ASec4} to extract fluid networks from the fluid data in these time frames. Each extracted fluid network characterizes vortical interactions in the flow field at a certain moment, where units are fluid elements and edge weights denote the induced velocities. We have filtered edges to exclude weak vortical interactions and control the effects of noises (see Appendix \ref{ASec4} for details).

\begin{figure}[!b]
\includegraphics[width=1\columnwidth]{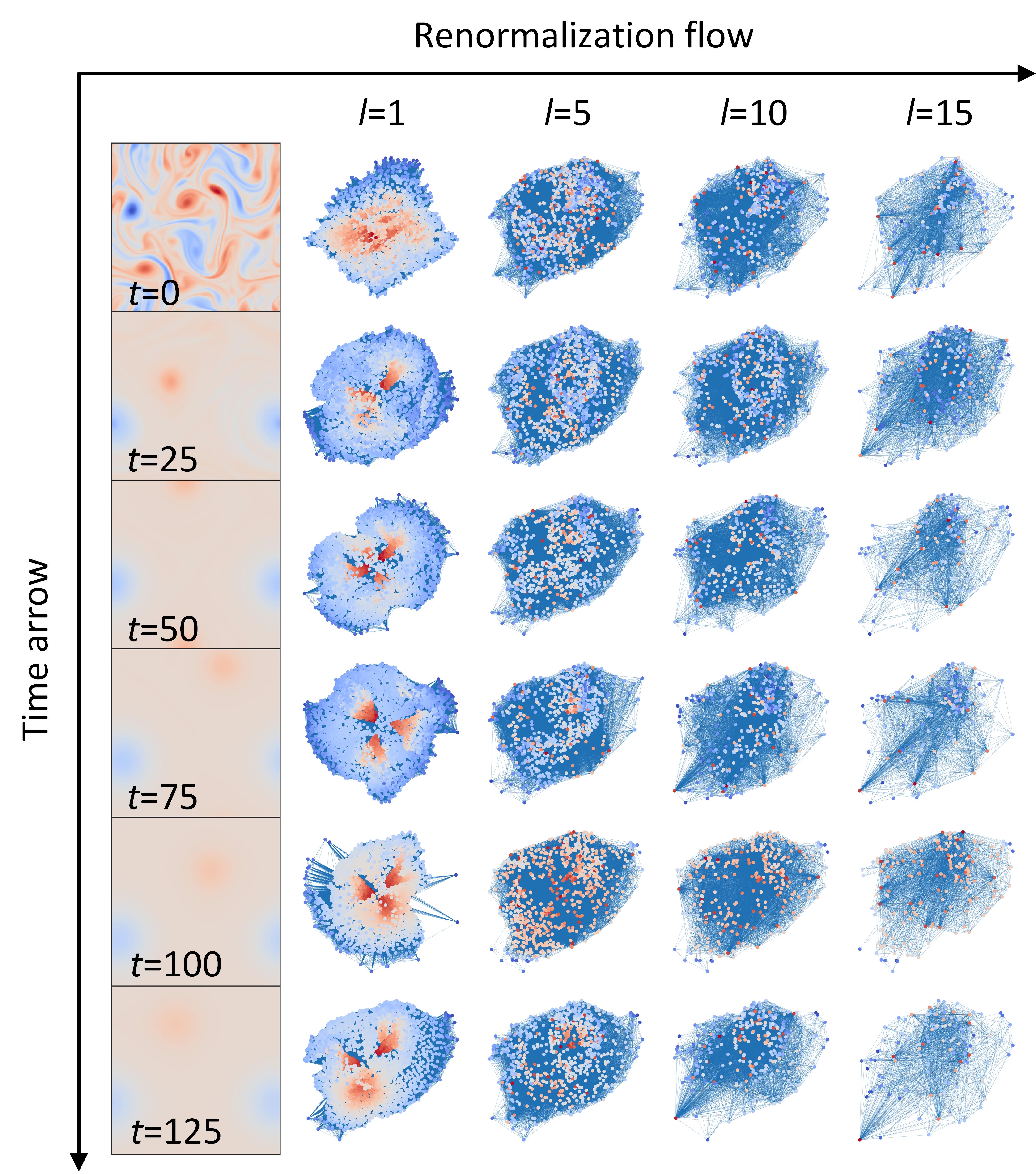}
\caption{\label{G0} Main pipeline of fluid data processing. Given a sequence of vortical fields across time ($t$), we extract their corresponding fluid networks and renormalize these fluid networks for multiple iterations ($l$). The colors of units in each network are determined using weighted degrees.} 
\end{figure}

After generating fluid flows and their network representations, we can simultaneously study their mechanics and network properties. On the one hand, we analyze the power-law decays of kinetic energy, $E$, and enstrophy, $\Omega$ in decaying turbulence \cite{boffetta2012two,kundu2015fluid}
\begin{align}
    E\propto t^{-\gamma},\;\;\;\;\;\;\;\;\;\Omega\propto t^{-\phi}, \label{EQ1}
\end{align}
where these physics quantities are calculated following Appendix \ref{ASec5} and fitted following Appendix \ref{ASec8}. Note that these power-law form decay laws do not apply to forced turbulence, where kinetic energy and enstrophy exhibit more sharp decays across time (see Fig. \ref{AG2} in Appendix \ref{ASec8} for instances). These decay processes convey information on how important fluid mechanics properties (e.g., the scales of energy change and fluid motion) evolve across time \cite{boffetta2012two,kundu2015fluid}.

On the other hand, we study spatial self-similarity \cite{taira2016network,meng2023scale,zhou2020power}, scale-invariance \cite{villegas2023laplacian,tian2024fast}, and control scaling \cite{klickstein2017energy,klickstein2018control}. For spatial self-similarity, we analyze it in its classic \cite{taira2016network} and generalized forms \cite{meng2023scale,zhou2020power}, which manifest as the power-law probability distributions of weighted degree, $\operatorname{deg}$, and degree-degree distance, $\eta$, respectively (see Appendix \ref{ASec6} for definitions). As demonstrated in Appendix \ref{ASec9}, these power-law behaviours can be equivalently and more efficiently verified by the power-law rank scaling of $\operatorname{deg}$ and $\eta$
\begin{align}
    \operatorname{deg}_{r}\propto r^{-\sigma},\;\;\;\;\;\;\;\;\;\eta_{r}\propto r^{-\chi}, \label{EQ2}
\end{align}
where $r$ denotes the rank after we sort these network quantities in a decreasing order (e.g., the $r$-th largest value among all samples is $\eta_{r}$). Compared with estimating power-law probability densities on empirical data, analyzing rank scaling is easier and more robust (i.e., one only need to sort the data). When necessary, we can also transform rank scaling exponents, $\sigma$ and $\chi$, to the power-law exponents of probability distributions via the simple algebra shown in Appendix \ref{ASec9}. Exponent $\sigma$ determines how the spatial patterns of vortical interactions (i.e., the induced velocities of vortex) are self-similar. Exponent $\chi$ describes how the structure imbalance of vortical interactions (i.e., a kind of generalized gradient of induced velocities) exhibits self-similarity. 

For scale-invariance, we use a renormalization group (RG) approach \cite{tian2024fast} introduced in Appendix \ref{ASec7} to realize the scale transformation of fluid networks. This framework, termed as random renormalization group, is a model-free RG developed for fast renormalizing the structure and dynamics of ultra-large systems \cite{tian2024fast}. By transforming fluid networks from fine-grained scales to coarse-grained scales, we analyze whether the macroscopic observables of fluid systems maintain invariant across scales. Specifically, we generate a renormalization flow, $\left(X^{\left(1\right)},\ldots,X^{\left(T\right)}\right)$, for each fluid network, where $X^{\left(1\right)}$ denotes the state of this network on the initial scale and $X^{\left(l\right)}$ is the transformed state on the $l$-th scale. As shown in Appendix \ref{ASec10}, the macroscopic observable of fluid network on the $l$-th scale can be defined as either the re-scaled ranked degree, $\bar{\operatorname{deg}}_{\bar{r}}^{\left(l\right)}$, or the re-scaled ranked degree-degree distance,  $\bar{\eta}_{\bar{r}}^{\left(l\right)}$. The evolution intensity of this macroscopic observable across scales can be measured as
\begin{align}
    \Delta= \big\langle \bar{\xi}_{\bar{r}}^{\left(l\right)}\big\rangle_{\bar{r},\;l},\label{EQ3}
\end{align}
where we define $\bar{\xi}_{\bar{r}}^{\left(l\right)}=\Big\vert\bar{\operatorname{deg}}_{\bar{r}}^{\left(l\right)}-\bar{\operatorname{deg}}_{\bar{r}}^{\left(1\right)}\Big\vert$ or $\bar{\xi}_{\bar{r}}^{\left(l\right)}=\Big\vert\bar{\eta}_{\bar{r}}^{\left(l\right)}-\bar{\eta}_{\bar{r}}^{\left(1\right)}\Big\vert$ to quantify the deviations of the macroscopic observable on the $l$-th scale from its initial state on the original scale. Notion $\langle\cdot\rangle_{\bar{r},\;l}$ denotes the averaging across all $\bar{r}$ and $l$. By measuring the evolution of network properties across scales, the value of $\Delta$ determines whether the fluid network satisfies scale-invariance. 

\begin{figure*}[!t]
\includegraphics[width=1\columnwidth]{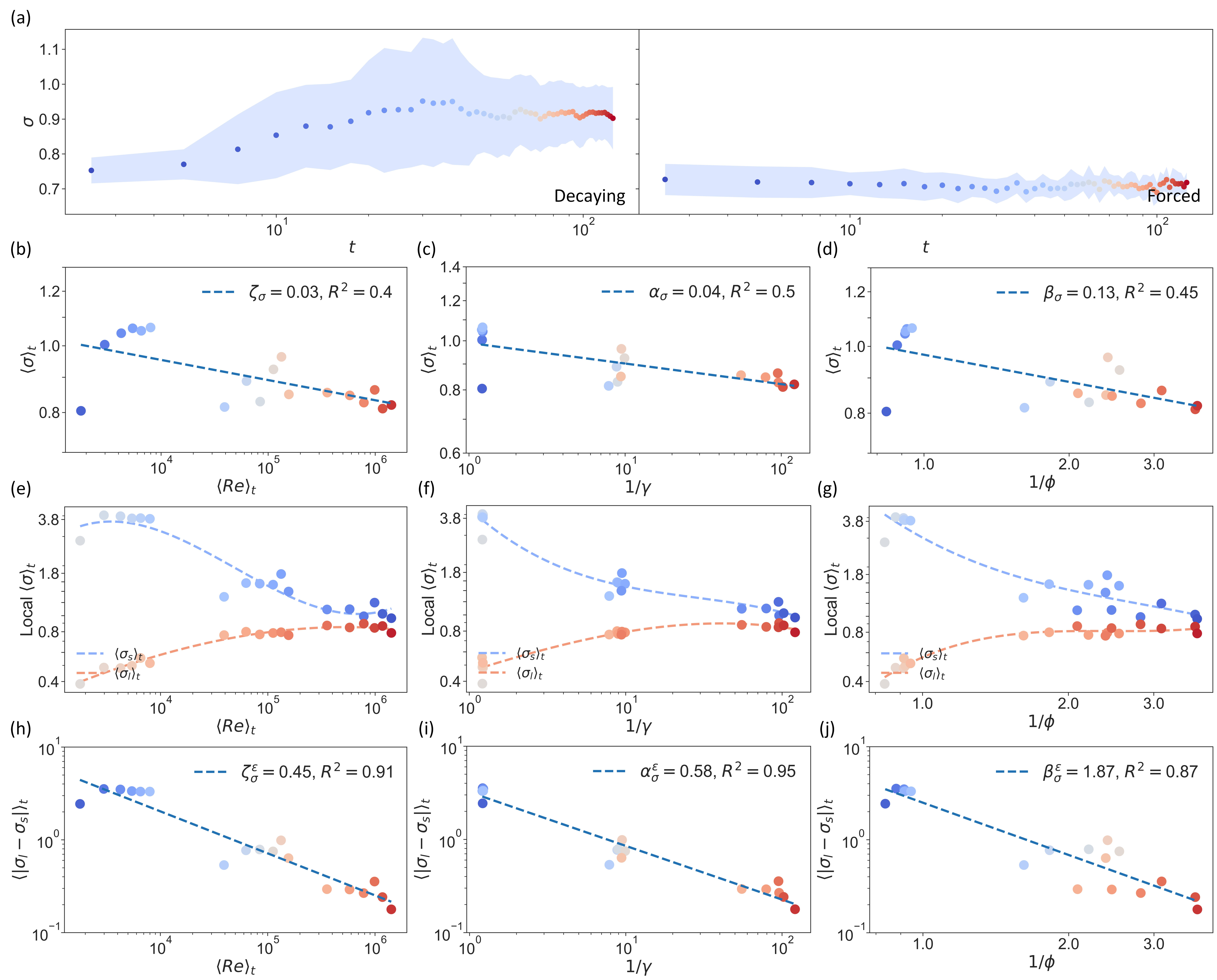}
\caption{\label{G1} Decay laws meet with classic spatial self-similarity. (a) illustrates the observations of the rank scaling exponent of weighted degrees, $\sigma$, in decaying and forced turbulence. In (a), scatters denote the mean value averaged across all combinations of $\left(\nu,u_{m}\right)$ and color areas represent the associated standard deviations. (b-d) respectively show how the time-averaged rank scaling exponent of weighted degrees, $\langle\sigma\rangle_{t}$, exhibits power-law scaling behaviours with respect to the time-averaged Reynolds number $\langle Re\rangle_{t}$, the inverse of kinetic energy decaying rate $\frac{1}{\gamma}$, and the inverse of enstrophy decaying rate $\frac{1}{\phi}$. (e-g) respectively illustrate how the local $\langle\sigma\rangle_{t}$ experiences bifurcations with respect to $\langle Re\rangle_{t}$, $\frac{1}{\gamma}$, and $\frac{1}{\phi}$. (h-j) respectively show the power-law scaling of $\langle \vert \sigma_{\text{l}}-\sigma_{\text{s}}\vert\rangle_{t}$ with respect to $\langle Re\rangle_{t}$, $\frac{1}{\gamma}$, and $\frac{1}{\phi}$. Scatters in (b-j) denote empirical data. Dashed lines in (b-d) and (h-j) represent fitted models while those in (e-g) stand for interpolated results. } 
\end{figure*}

For control scaling, we consider a control problem with single input and single control target on fluid networks in Appendix \ref{ASec6}. This problem, as shown in Appendix \ref{ASec6}, requires a minimum cost, $E_{c}$, to realize a non-trivial control (here a trivial control refers to the control process without any effect) \cite{klickstein2017energy,klickstein2018control}. Applying the infinite path network approximation, we can approximate $E_{c}$ as a function of $\ell$, the distance between input and control target in a fluid network, following Ref. \cite{klickstein2018control}. In Appendix \ref{ASec11}, we show an exponential scaling of the approximated minimum control cost 
\begin{align}
E_{c}\propto \exp\left(\kappa \ell\right),\label{EQ4}
\end{align}
which describes how the control cost experiences explosive increase as the control path in a given fluid network elongates.

After quantifying all these mechanics and network properties, we can explore their joint relations.

\section{Decay laws and classic spatial self-similarity}\label{Sec3}
In this section, we explore how the decays of kinetic energy and enstrophy affect the classic spatial self-similarity property \cite{taira2016network} of fluid networks. After measuring the rank scaling exponent of weighted degrees, $\sigma$, across different conditions of $\left(\nu,u_{m}\right)$ and different time frames, we first calculate the means and  standard deviations of $\sigma$ across all combinations of $\left(\nu,u_{m}\right)$ at each moment $t$. As shown in Fig. \ref{G1}(a), compared with forced turbulence, decaying turbulence has a more variable power-law exponent $\sigma$ across different conditions of $\left(\nu,u_{m}\right)$ because there exist larger standard deviations of $\sigma$ in all time frames. Placing this observation in the context of fluid mechanics, we can know that the absence of external forces (e.g., the classic Kolmogorov forcing \cite{rollin2011variations}) provides opportunities for fluid properties (e.g., viscosity and velocity bound) to freely regulate the characteristics of fluid networks.

Specifically, the regulation effects on the classic spatial self-similarity of the fluid networks associated with decaying turbulence are summarized in Figs. \ref{G1}(b-j). To focus on the static relations between fluid properties and network characteristics, we exclude the time dimension by considering the time-averaged rank scaling exponent of weighted degrees, $\langle\sigma\rangle_{t}$, the time-averaged Reynolds number, $\langle Re\rangle_{t}$, the kinetic energy decaying rate, $\gamma$, and the enstrophy decaying rate, $\phi$, in our analysis. A direct observation in Figs. \ref{G1}(b-d) is that $\langle\sigma\rangle_{t}$ approximately exhibits power-law scaling behaviours with respect to fluid properties or their inverse values, i.e., $\langle\sigma\rangle_{t}\propto \langle Re\rangle_{t}^{-\zeta_{\sigma}}$, $
\langle\sigma\rangle_{t}\propto \left(1/\gamma\right)^{-\alpha_{\sigma}}$, and $\langle\sigma\rangle_{t}\propto \left(1/\phi\right)^{-\beta_{\sigma}}$. The internal consistency among these scaling behaviours is comprehensible since $\gamma$ and $\phi$, as suggested in Appendix \ref{ASec8}, are negatively correlated with the Reynolds number \cite{chasnov1997decay}. The actual values of power-law exponents, $\zeta_{\sigma}$, $\alpha_{\sigma}$, and $\beta_{\sigma}$, should be treated carefully because we observe non-negligible errors in model fitting in Figs. \ref{G1}(b-d) (i.e., with an acceptable but not sufficiently high accuracy of $R^{2}\in\left[0.4,0.6\right]$).

\begin{figure*}[!t]
\includegraphics[width=1\columnwidth]{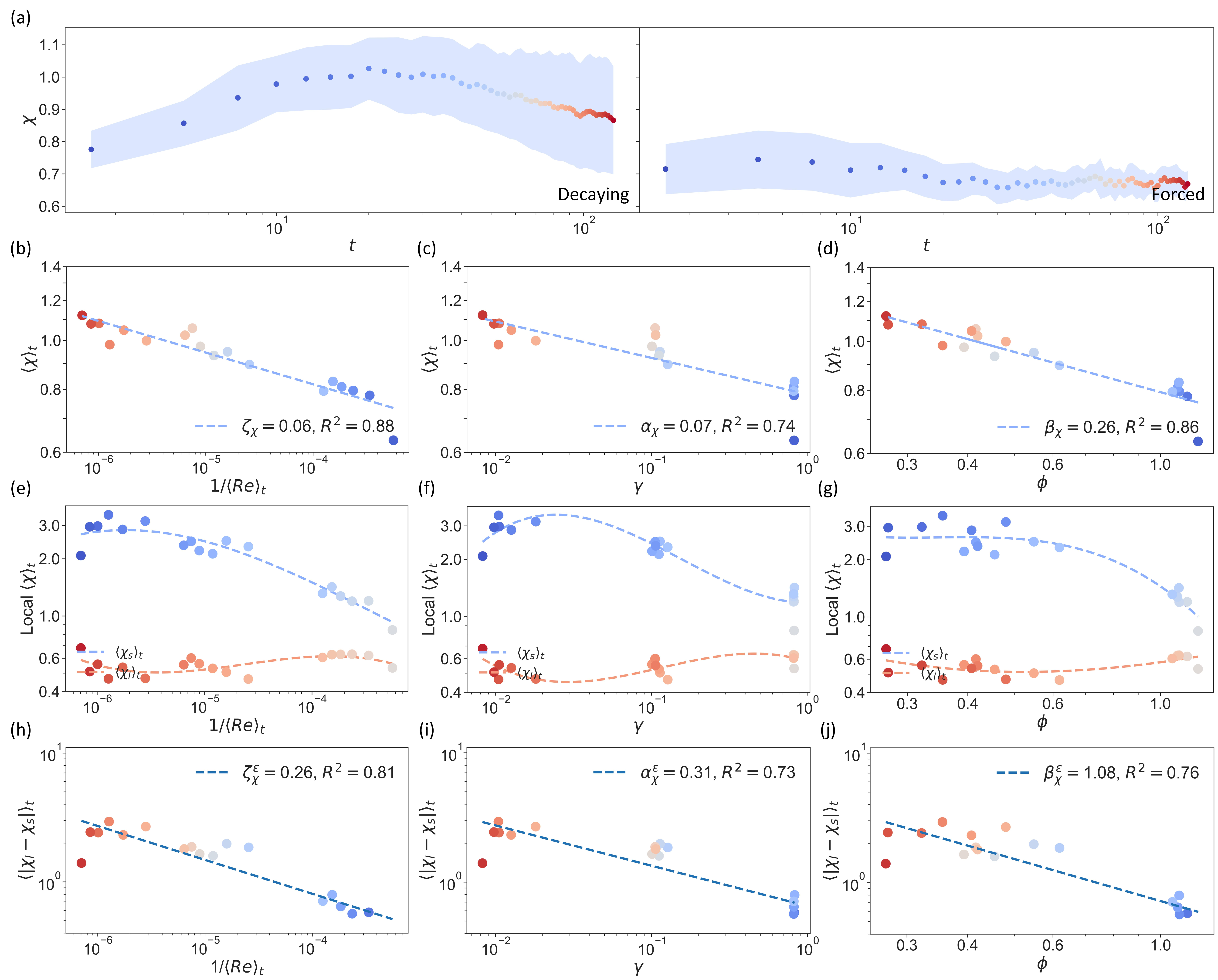}
\caption{\label{G2} Decay laws meet with generalized spatial self-similarity. (a) illustrates the observations of the rank scaling exponent of degree-degree distances, $\chi$, in decaying and forced turbulence. Scatters denote the mean value averaged across all combinations of $\left(\nu,u_{m}\right)$ and color areas represent the associated standard deviations. (b-d) respectively show how the time-averaged rank scaling exponent of degree-degree distance, $\langle\chi\rangle_{t}$, exhibits power-law scaling behaviours with respect to the inverse of time-averaged Reynolds number $\frac{1}{\langle Re\rangle_{t}}$, the kinetic energy decaying rate $\gamma$, and the enstrophy decaying rate $\phi$. (e-g) respectively illustrate how the local $\langle\chi\rangle_{t}$ experiences bifurcations with respect to $\frac{1}{\langle Re\rangle_{t}}$, $\gamma$, and $\phi$. (h-j) respectively show the power-law scaling of $\langle \vert \chi_{\text{l}}-\chi_{\text{s}}\vert\rangle_{t}$ with respect to $\frac{1}{\langle Re\rangle_{t}}$, $\gamma$, and $\phi$. Scatters in (b-j) denote empirical data. Dashed lines in (b-d) and (h-j) represent fitted models while those in (e-g) stand for interpolated results. } 
\end{figure*}

How do these errors arise? The suspicion of the power-law exponent $\sigma$ of weighted degrees can not be eliminated because there is no theoretical guarantee that weighted degrees must exhibit power-law rank scaling (i.e., errors may occur in $\sigma$ when we estimate a power-law model from non-power-law data). This idea motivates us to verify whether fluid networks deviate from spatially self-similar states (i.e., the power-law rank scaling breaks down). Because $\langle Re\rangle_{t}$, $\gamma$, and $\phi$ are related to the vanishing rate of small vortex in turbulence, we decide to subdivide Eq. (\ref{EQ2}) into two sub-intervals, which respectively correspond to large and small vortex
\begin{align}
\operatorname{deg}_{r}\propto r^{-\sigma_{\text{l}}},\;r<0.5M,\label{EQ5}\\
\operatorname{deg}_{r}\propto r^{-\sigma_{\text{s}}},\;r\geq 0.5M.\label{EQ6}
\end{align}
 Here $\sigma_{\text{l}}$ and $\sigma_{\text{s}}$ denote the local power-law exponents associated with large and small vortex. If the weighted degrees of a fluid network truly exhibits the power-law rank scaling, we expect to see $\sigma_{\text{l}}\simeq \sigma_{\text{s}}$ (i.e., all parts of a power-law model share a common slope in the log-log plot). As shown in Figs. \ref{G1}(e-g), two local power-law exponents, $\sigma_{\text{l}}$ and $\sigma_{\text{s}}$, gradually converge to each other as $\langle Re\rangle_{t}$ increases (or equivalently, as $\gamma$ and $\phi$ decrease). In other words, as the fluid flow becomes increasingly turbulent and the decays of kinetic energy and enstrophy become much slower, small vortex become more enduring in the flow to maintain the classic spatial self-similarity. In an opposite direction (i.e., $\langle Re\rangle_{t}$ decreases), the accelerated vanishing of small vortex contributes to the bifurcations of $\sigma_{\text{l}}$ and $\sigma_{\text{s}}$ in Figs. \ref{G1}(e-g) as well as the breaking of the classic spatial self-similarity.

In Figs. \ref{G1}(h-j), we can observe the power-law scaling of the absolute difference between $\sigma_{\text{l}}$ and $\sigma_{\text{s}}$ with respect to fluid properties or their inverse values

\begin{align}
\langle\vert\sigma_{\text{l}}-\sigma_{\text{s}}\vert\rangle_{t}&\propto \langle Re\rangle_{t}^{-\zeta_{\sigma}^{\varepsilon}},\label{EQ7}\\
\langle\vert\sigma_{\text{l}}-\sigma_{\text{s}}\vert\rangle_{t}&\propto \left(\frac{1}{\gamma}\right)^{-\alpha_{\sigma}^{\varepsilon}},\label{EQ8}\\
\langle\vert\sigma_{\text{l}}-\sigma_{\text{s}}\vert\rangle_{t}&\propto \left(\frac{1}{\phi}\right)^{-\beta_{\sigma}^{\varepsilon}}.\label{EQ9}
\end{align}
These power-law scaling behaviours hold with a high accuracy, indicating how the fluid properties of decaying turbulence determine the existence of classic spatial self-similarity in fluid networks.

\section{Decay laws and generalized spatial self-similarity}\label{Sec4}
The analysis paradigm presented above can be directly extended to the generalized spatial self-similarity, i.e., the spatial self-similarity behaviours of connectivity imbalance \cite{meng2023scale,zhou2020power} in fluid networks.

As shown in Fig. \ref{G2}(a), the power-law exponent $\chi$ of degree-degree distances, similar to exponent $\sigma$, has higher variability across different conditions of $\left(\nu,u_{m}\right)$ in decaying turbulence. The absence of external forces lays the key foundation for the generalized spatial self-similarity to be regulated by fluid properties, motivating us to primarily focus on decaying turbulence. 

In Figs. \ref{G2}(b-d), we observe clear power-law scaling behaviours of the time-averaged scaling exponent of connectivity imbalance, $\langle\chi\rangle_{t}$, with respect to fluid properties or their inverse values, i.e., $\langle\chi\rangle_{t}\propto \left(1/\langle Re\rangle_{t}\right)^{-\zeta_{\chi}}$, $\langle\chi\rangle_{t}\propto \gamma^{-\alpha_{\chi}}$, and $\langle\chi\rangle_{t}\propto \phi^{-\beta_{\chi}}$. We discover that $\langle\chi\rangle_{t}$ evolves in an opposite direction of $\langle\sigma\rangle_{t}$. Then, we verify whether degree-degree distances truly exhibit power-law rank scaling under different conditions of fluid properties, where we analyze the numerical convergence of $\chi_{\text{l}}$ and $\chi_{\text{s}}$, two local power-law exponents respectively associated with large and small vortex. As Figs. \ref{G2}(e-g) suggest, exponents $\langle\chi_{\text{l}}\rangle_{t}$ and $\langle\chi_{\text{s}}\rangle_{t}$ tend to converge to each other when $\langle Re\rangle_{t}$ decreases (or equivalently, when $\gamma$ and $\phi$ increase). The convergence can be characterized by the following power-law scaling relations
\begin{align}
\langle\vert\chi_{\text{l}}-\chi_{\text{s}}\vert\rangle_{t}&\propto \left(\frac{1}{\langle Re\rangle_{t}}\right)^{-\zeta_{\chi}^{\varepsilon}},\label{EQ10}\\
\langle\vert\chi_{\text{l}}-\chi_{\text{s}}\vert\rangle_{t}&\propto \gamma^{-\alpha_{\chi}^{\varepsilon}},\label{EQ11}\\
\langle\vert\chi_{\text{l}}-\chi_{\text{s}}\vert\rangle_{t}&\propto \phi^{-\beta_{\chi}^{\varepsilon}},\label{EQ12}
\end{align}
which are shown in Figs. \ref{G2}(h-j) with high accuracy. 

Taken together, what we can conclude from Figs. \ref{G2}(b-j) is the indispensable role of turbulent extent in maintaining the spatial self-similarity of connectivity imbalance. In the case where freely decaying fluid flows are less turbulent (i.e., when $\langle Re\rangle_{t}$ is small), larger vortex are possible to emerge and small vortex vanish with higher decay rates, creating larger variability in the connectivity imbalance of vortical interactions and making it possible for connectivity imbalance to exhibit spatial self-similar behaviours. When freely decaying fluid flows become more turbulent, fluid quantity decays become slower and there lacks a sufficient driving factor to ensure the variability of connectivity imbalance (in our analysis, we observe that the time-averaged coefficient of variation \cite{abdi2010coefficient} of $\eta$ decreases from $1.825$ to $0.924$ as $\langle Re\rangle_{t}$ increases), leading to the breaking of spatial self-similarity in connectivity imbalance. Please note that Eqs. (\ref{EQ10}-\ref{EQ12}) do not conflict with Eqs. (\ref{EQ7}-\ref{EQ9}) because the imbalance of vortical interactions characterized by $\eta$, a kind of discrete gradient of interaction strengths, does not necessarily evolve with interaction strengths in the same direction.

\begin{figure*}[!t]
\includegraphics[width=1\columnwidth]{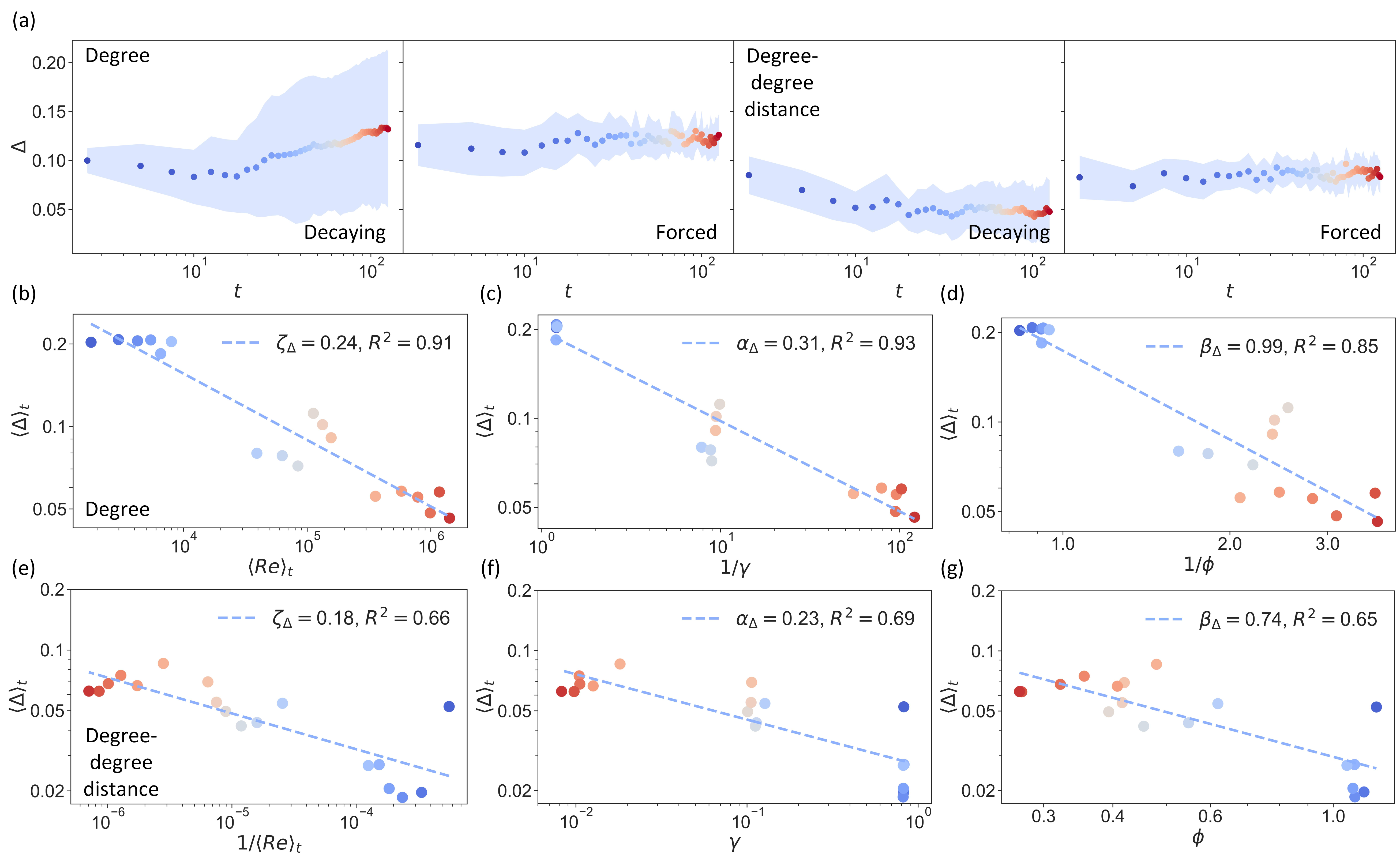}
\caption{\label{G3} Decay laws meet with scale-invariance. (a) illustrates $\Delta$, the evolution intensities of different macroscopic observables, across time, where scatters denote the mean value averaged across all combinations of $\left(\nu,u_{m}\right)$ and color areas represent the associated standard deviations. (b-d) respectively show the power-law scaling of the time-averaged evolution intensity, $\langle\Delta\rangle_{t}$, of weighted degrees with respect to the time-averaged Reynolds number $\langle Re\rangle_{t}$, the inverse of kinetic energy decaying rate $\frac{1}{\gamma}$, and the inverse of enstrophy decaying rate $\frac{1}{\phi}$. (e-g) respectively illustrate the power-law scaling of the time-averaged evolution intensity, $\langle\Delta\rangle_{t}$, of degree-degree distances with respect to the inverse of time-averaged Reynolds number $\frac{1}{\langle Re\rangle_{t}}$, the kinetic energy decaying rate $\gamma$, and the enstrophy decaying rate $\phi$. In (b-g), scatters denote empirical data while dashed lines represent fitted models.} 
\end{figure*}

\begin{figure*}[!t]
\includegraphics[width=1\columnwidth]{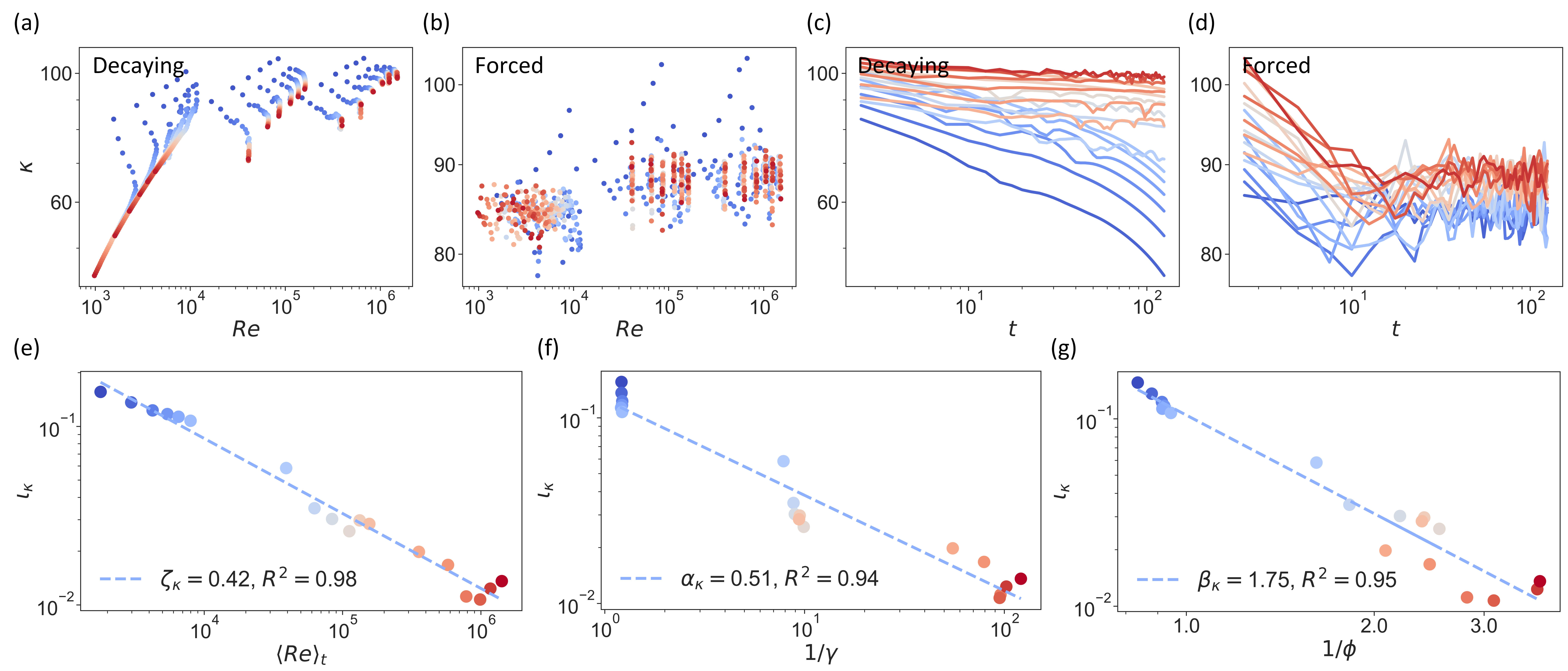}
\caption{\label{G4} Decay laws meet with control scaling. (a-b) respectively show the control scaling exponents, $\kappa$, of decaying and forced turbulence across different conditions of $\left(\nu,u_{m}\right)$, where the colors of scatters, changing from blue to red, denote the increase of time steps, $t$. (c-d) respectively illustrate the control scaling exponents, $\kappa$, of decaying and forced turbulence across time, where line colors, changing from blue to red, represent different combinations of $\left(\nu,u_{m}\right)$. It can be seen that $\kappa$ exhibits power-law scaling with respect to time $t$ in decaying turbulence, where the scaling exponent is denoted by $\iota_{\kappa}$ in our analysis. (e-g) respectively show the power-law scaling of $\iota_{\kappa}$ with respect to the time-averaged Reynolds number $\langle Re\rangle_{t}$, the inverse of kinetic energy decaying rate $\frac{1}{\gamma}$, and the inverse of enstrophy decaying rate $\frac{1}{\phi}$.} 
\end{figure*}

\section{Decay laws and scale-invariance}\label{Sec5}
Our analysis can be further utilized to explore how fluid properties interact with the scale-invariance of fluid networks. The key difference between spatial self-similarity and scale-invariance is that spatial self-similarity refers to the similar pattern shared by different local regions (e.g., shared by a region and its sub-region) of fluid networks on a given scale while scale-invariance refers to the invariant property of fluid networks across different scales (see Appendix \ref{ASec2} for details).

Similar to the situations in Figs. \ref{G1}-\ref{G2}, the freely decaying condition without any external disturbance enables fluid properties to regulate the scale-invariance of fluid networks effectively (see Fig. \ref{G3}(a)). If we analyze weighted degrees across scales, the power-law scaling behaviours of their time-averaged evolution
intensity can be observed in Figs. \ref{G3}(b-d), i.e., $\langle\Delta\rangle_{t}\propto \langle Re\rangle_{t}^{-\zeta_{\Delta}}$, $\langle\Delta\rangle_{t}\propto \left(1/\gamma\right)^{-\alpha_{\Delta}}$, $\langle\Delta\rangle_{t}\propto \left(1/\phi\right)^{-\beta_{\Delta}}$. These scaling behaviours suggest that vortical interactions tend to be scale-invariant as decaying turbulence becomes more turbulent. If we study degree-degree distance across scales, we can also observe the power-law scaling of the time-averaged evolution intensity in Figs. \ref{G3}(e-g), i.e., $\langle\Delta\rangle_{t}\propto \left(1/\langle Re\rangle_{t}\right)^{-\zeta_{\Delta}}$, $\langle\Delta\rangle_{t}\propto \gamma^{-\alpha_{\Delta}}$, $\langle\Delta\rangle_{t}\propto \phi^{-\beta_{\Delta}}$, suggesting that the imbalance of vortical interactions becomes more scale-invariant when the flow is less turbulent. These two types of scaling behaviours consistently imply that the variability of macroscopic observables regulated by fluid decay laws is indispensable in guaranteeing scale-invariance. The variability of vortical interactions increases with the diversities of vortex and favours slower kinetic energy and enstrophy decays (the time-averaged coefficient of variation \cite{abdi2010coefficient} of $\operatorname{deg}$ increases from $0.779$ to $1.738$ as $\langle Re\rangle_{t}$ enlarges). The variability of the imbalance of vortical interactions depends on the decays of fluid quantities and favours significant decays (as we have reported previously, the time-averaged coefficient of variation \cite{abdi2010coefficient} of $\eta$ changes from $0.924$ to $1.825$ as $\langle Re\rangle_{t}$ decreases). Combining our observations in Figs. \ref{G3}(b-g) with Eqs. (\ref{EQ7}-\ref{EQ12}), what we can further derive is
\begin{align}
\log\langle\Delta\rangle_{t}&\propto\log\langle\vert\sigma_{\text{l}}-\sigma_{\text{s}}\vert\rangle_{t}\label{EQ13}
\end{align} 
or 
\begin{align}
\log\langle\Delta\rangle_{t}&\propto\log\langle\vert\chi_{\text{l}}-\chi_{\text{s}}\vert\rangle_{t}.\label{EQ14}
\end{align}
These two relations indicate that vortical interactions, or their connectivity imbalance distributions, in decaying turbulence tend to be scale-invariant when they become spatially self-similar on the original scale.

Moreover, Eqs. (\ref{EQ13}-\ref{EQ14}) also relate between scale-transformation and box-covering in the fluid networks of decaying turbulence. Because scale-transformation in our work is realized by coarse graining the vortical elements with strong correlations into macro-units (see Appendix \ref{ASec7}), we can treat each macro-unit as a box whose size equals the number of contained vortical elements. When scale-transformation and box-covering become consistent, their direct corollaries make fluid networks satisfy two properties simultaneously: (1) the average size of macro-units approaches to the $d_{h}$-th root of macro-unit number when a fluid network become spatially self-similar, where $d_{h}$ denotes the Hausdorff dimension \cite{falconer2004fractal} of this fluid network; (2) the network of boxes share similar topological properties with the original fluid network. Although we do not focus on these two properties in this study, they might be useful in the network analysis  of fluids in future works.

\section{Decay laws and control scaling}\label{Sec6}
Finally, we explore the effects of fluid decay laws on control scaling \cite{klickstein2017energy,klickstein2018control}. In our analysis, we focus on the exponential scaling behaviour, $E_{c}\propto \exp\left(\kappa \ell\right)$, of the minimum control cost, $E_{c}$, with respect to the distance between input and control target in a fluid network, $\ell$.

In Figs. \ref{G4}(a-b), we show the evolution of scaling exponent, $\kappa$, across different time frames and conditions of $\left(\nu,u_{m}\right)$. Different from the irregular situations in forced turbulence (Fig. \ref{G4}(b)), exponent $\kappa$ experiences clear reductions across time in the decaying turbulence defined with every combination of $\left(\nu,u_{m}\right)$ (Fig. \ref{G4}(a)). Such kind of temporal reduction is presented in Figs. \ref{G4}(c-d), where we can observe the power-law scaling behaviours of $\kappa$ with respect to time $t$ in decaying turbulence (Fig. \ref{G4}(c))
\begin{align}
    \kappa\propto t^{-\iota_{\kappa}}.\label{EQ15}
\end{align}
According to Eq. (\ref{EQ15}), the growth speed of control cost $E_{c}$ with respect to control distance $\ell$ decreases with time. Driven by kinetic energy and enstrophy decays, controlling decaying turbulence becomes more economic as time passes (i.e., the growth of $E_{c}$ becomes slower). A similar phenomenon can be seen in forced turbulence except that the temporal reduction of $\kappa$ does not follow a power-law model (Fig. \ref{G4}(d)). Consequently, we primarily focus on decaying turbulence in the subsequent analysis.

What are the roles of fluid properties in regulating the temporal reduction of $\kappa$? In Figs. \ref{G4}(e-g), we observe the power-law scaling behaviours of exponent $\iota_{\kappa}$ with respect to fluid properties or their inverse values
\begin{align}
\iota_{\kappa}&\propto \langle Re\rangle_{t}^{-\zeta_{\kappa}^{\varepsilon}},\label{EQ16}\\
\iota_{\kappa}&\propto \left(\frac{1}{\gamma}\right)^{-\alpha_{\kappa}^{\varepsilon}},\label{EQ17}\\
\iota_{\kappa}&\propto \left(\frac{1}{\phi}\right)^{-\beta_{\kappa}^{\varepsilon}}.\label{EQ18}
\end{align}
As suggested by these scaling behaviours, the temporal reduction rate of exponent $\kappa$ decreases when the fluid flow becomes increasingly turbulent. When $\langle Re\rangle_{t}$ is sufficiently large, the temporal reduction of $\kappa$ becomes negligible and the minimum cost of controlling fluid networks is generally time-invariant.

How are these findings related to fluid mechanics? In fact, although the above analysis is implemented using the terminology of controlling, a control action can also be understand as an interaction, i.e., how do the perturbations on one vortical element affect another vortical element separated by a given distance? The minimum control cost reflects the minimum perturbation strength required for realizing a non-negligible propagation of perturbation effects. Therefore, the regulations on control scaling by fluid properties can also be interpreted as the decisive effects of fluid properties on vortical interactions over different spatial ranges.

\section{Conclusion}

To conclude, we have revealed four kinds of fluid-network relations in this work, which quantify how fundamental fluid properties, including kinetic energy and enstrophy decay laws \cite{boffetta2012two,kundu2015fluid}, interact with defining network characteristics, such as spatial self-similarity \cite{taira2016network,meng2023scale,zhou2020power}, scale-invariance \cite{villegas2023laplacian,tian2024fast}, and control scaling \cite{klickstein2017energy,klickstein2018control}. Although these concepts seem to be irrelevant and lack clear connections, we suggest the existences of robust power-law scaling relations among them, which quantify how fluid properties are reflected in network characteristics  or vice versa. For a brief summary, we list our main findings below.
\begin{itemize}
    \item[(1) ] The existence of classic spatial self-similarity \cite{taira2016network} in fluid networks critically relays on the slow decays of fluid quantities. When fluid flows are increasingly turbulent and there is no external force, vortex become more enduring to ensure a higher diversity among vortex (i.e., large vortex coexist with small vortex) and enable vortical interactions to exhibit power-law behaviours on the spatial domain. Quantitatively, the deviation degrees of fluid networks from classic self-similar states satisfy power-law scaling relations with the time-averaged Reynolds number, the inverse of kinetic energy decaying rate, and the inverse of enstrophy decaying rate in Eqs. (\ref{EQ7}-\ref{EQ9}), which serve as the first group of fluid-network relations.       
    
    \item[(2) ]As described by the second group of fluid-network relations in Eqs. (\ref{EQ10}-\ref{EQ12}), the generalized spatial self-similarity \cite{meng2023scale,zhou2020power} of fluid networks more favours the non-negligible decays of fluid quantities. As freely decaying fluid flows become less turbulent, the decays of kinetic energy and enstrophy are observable on small time scales, which imply higher variability of the spatial imbalance extents of vortical interactions to enable them self-organize to spatially self-similar states.
    \item[(3) ]The third group of fluid-network relations suggest that the departure degrees of fluid networks from being scale-invariant \cite{villegas2023laplacian,tian2024fast} share similar power-law scaling behaviours with the deviation degrees of spatial self-similarity. Fluid properties can be used as intermediary variables to obtain the proportional relation between scale-invariance and spatial self-similarity in fluid networks (see Eqs. (\ref{EQ13}-\ref{EQ14}) for details). At least for vortical interactions and their spatial imbalance, being invariant across different scales is closely related to being spatially self-similar on each scale.
\item[(4) ]In the fourth group of fluid-network relations, Eqs. (\ref{EQ15}-\ref{EQ18}), the minimum cost \cite{klickstein2017energy,klickstein2018control} of controlling fluid networks over a given range has a growth rate $\kappa$, which exhibits power-law decays across time just like kinetic energy and enstrophy. The temporal decay speed of $\kappa$ follows power-law scaling relations with fluid properties. As fluid flows get more turbulent, the temporal decay speed of $\kappa$ become sufficiently small such that the minimum control cost tends to be time-invariant. In other words, the energy required to affect one vortical element by perturbing another vortical element over a given distance does not change across time in sufficiently turbulent fluid flows.

\end{itemize}

Hopefully, these discovered fluid-network relations may serve as quantitative bridges between fluid mechanics \cite{boffetta2012two,kundu2015fluid} and complex network theories \cite{costa2011analyzing}. Because our analysis is mainly implemented on two-dimensional turbulent flows (i.e., freely decaying and forced turbulence), we suggest the necessity to verify or generalize these relations in other kinds of fluid flows of higher dimensions. Meanwhile, we propose to explore the theoretical foundations of our results since the present fluid-network relations are derived statistically. Why is such an exploration worthwhile? Perhaps one of the appropriate reasons is that fluid systems with heterogeneous internal interactions are highly non-trivial in the aspect of mechanics while network theories are essentially effective in analyzing heterogeneity. Although the anisotropy of fluid flows \cite{biferale2005anisotropy} is not included in our present research, studying it using network analysis will be one of the ultimate objectives in future works.

\section*{Acknowledgements}This project is supported by the Artificial and General Intelligence Research Program of Guo Qiang Research Institute at Tsinghua University (2020GQG1017) as well as the Tsinghua University Initiative Scientific Research Program. 

\appendix
\section{Mini review of network structure identification}\label{ASec1}

To make our analysis comprehensible for interdisciplinary researchers, we present a focused review on the common techniques for identifying the implicit network structures of fluids, which establish mappings from fluids to networks. 

\paragraph*{Eulerian perspective.} From an Eulerian perspective, a representative method is to define units as vortical elements and characterize edges according to induced vorticities \cite{nair2015network,taira2016network}. The vorticity field in this method can also be replaced by the acoustic power field to analyze turbulent combustors \cite{krishnan2019emergence,kawano2023complex}. Another practical approach is to extract networks from the similarities (e.g., correlations) among physical quantities (e.g., kinetic energies and velocities) on different locations. Similarities are filtered via a threshold such that only strong correlations can remain and define edges to represent inter-subsystem interactions \cite{scarsoglio2016complex,iacobello2018spatial,tandon2023multilayer}. This approach is also popular outside the field of fluid mechanics \cite{donges2009complex,zhou2015teleconnection,tupikina2016correlation}. As an alternative, the third way is to construct weighted Gabriel networks based on proximity strengths, which sports to classify vortical flows \cite{krueger2019quantitative}.

\paragraph*{Lagrangian perspective.} Apart from the Eulerian perspective, one can also build networks using particle trajectories from a Lagrangian viewpoint. For instance, vortical networks \cite{nair2015network,taira2016network} can be generalized into a Lagrangian form by tracking the movement of vortical community centroids \cite{meena2018network}, similarity networks \cite{scarsoglio2016complex,iacobello2018spatial,tandon2023multilayer} can be defined on particle trajectories (i.e., tracers) \cite{schlueter2019model}, and proximity networks can be constructed on fluid particles to quantify the wall-normal turbulent mixing intensity \cite{iacobello2019lagrangian,perrone2020wall} or the effects of the Reynolds number \cite{perrone2021network}.

\paragraph*{Time series view.} Furthermore, one can choose to extract visibility \cite{murugesan2015combustion,guan2023multifractality} and transition \cite{kaiser2014cluster} networks purely from the time series of fluid data \cite{iacobello2021review}, which serves as an independent way different from the Eulerian and the Lagrangian setups. 

\section{Concept clarification}\label{ASec2}
Because our work attempts to bridge between network science and fluid mechanics, we inevitably deal with some concepts with multiple meanings in these two fields. To avoid confusions, we clarify the definitions of these concepts used in our analysis.

\paragraph*{Spatial self-similarity.} The spatial self-similarity of complex networks is characterized by the power-law behaviours of degree distributions or other related macroscopic observables. Such kind of self-similarity is embodied in the consistent topological properties shared by a complex network and its sub-networks on a given scale \cite{barabasi2009scale}.

\paragraph*{Scale-invariance.} We denote scale-invariance as the strictly invariant properties of networks under the scale transformation (e.g., by renormalization groups). Although scale-invariance and spatial self-similarity coincide with each other on occasions, many scale-invariant structures (e.g., lattices where all units share the same degree) do not exhibit power-law behaviours and many self-similar structures (e.g., the Barab{\'a}si-Albert network \cite{albert2002statistical}) are conditionally invariant under scale transformations (e.g., feature two or more invariant scale regions) \cite{villegas2023laplacian,tian2024fast}. Moreover, a coarse graining procedure does not necessarily transform a network into its sub-networks, making the analyses of scale-invariance and spatial self-similarity non-equivalent in some situations.

\paragraph*{Scaling behaviours.} Scaling behaviours are closely related to scale transformations \cite{lovejoy2022scaling}. A network exhibits scaling behaviours if one of its macroscopic observable follows a deterministic function of scales characterized by fixed scaling exponents. Common scaling behaviours include exponential and power-law scaling.

\paragraph*{Decay laws.} Although decay laws are widespread in diverse systems, here we only use decay laws to denote the decreasing trends of fluid properties (e.g., kinetic energy and enstrophy) with time, spatial distance, or other control parameters \cite{skrbek2000decay}.

\begin{figure*}[!t]
\includegraphics[width=1\columnwidth]{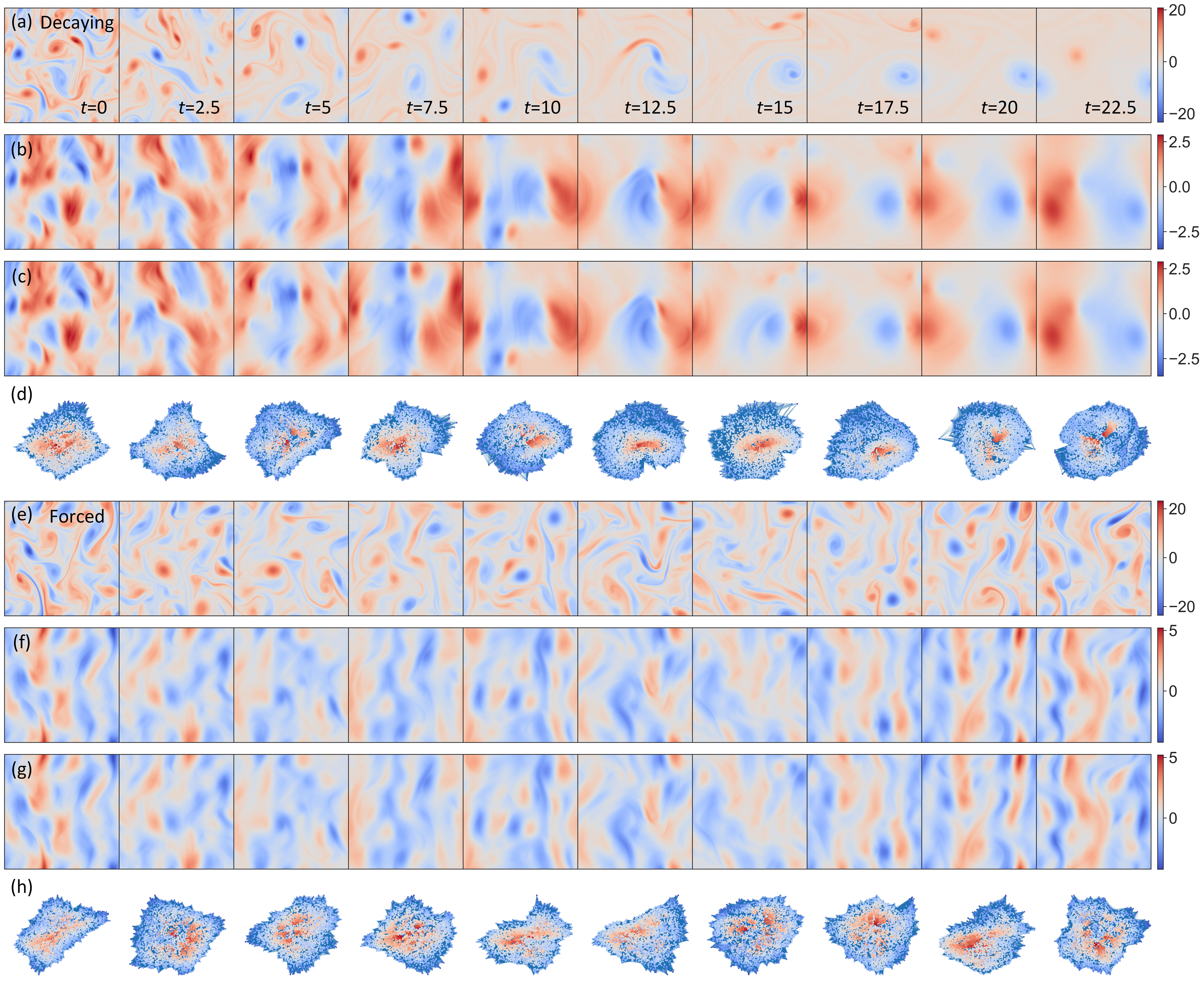}
\caption{\label{AG0} Two instances of fluid and network generation. (a) The generated sequences of vorticity fields are illustrated, where we visualize the first ten time frames as examples. (b) The associated velocity fields (first component) are presented. (c) The associated velocity fields (second component) are illustrated. (d) The fluid networks extracted from vorticity fields are shown, where node colors are defined according to weighted degrees. All results in (a-d) are extracted from a decaying turbulent flow with $\nu=10^{-3}$ and $u_{m}=4$. (e-h) Meanwhile, the results derived from a forced turbulent flow with $\nu=10^{-3}$ and $u_{m}=4$ are shown for comparison. } 
\end{figure*}

\section{Generation of turbulent flows}\label{ASec3}
In our work, we utilize the JAX-CFD \cite{Kochkov2021-ML-CFD,Dresdner2022-Spectral-ML}, a computational fluid dynamics framework in JAX \cite{jax2018github}, to simulate two-dimensional turbulent flows. Specifically, we define three viscosity conditions, $\nu\in\{10^{-2}, 10^{-3},10^{-4}\}$, and six velocity bound conditions, $u_{m}\in\{4, 6,\ldots,12,14\}$, to simulate decaying and forced turbulence on a spatial domain $D=\left[0,2\pi\right]\times \left[0,2\pi\right]$ during an interval of $125$ seconds. Domain $D$ is discretized into $256\times 256$ Arakawa grids \cite{arakawa1977computational}.

The corresponding incompressible Navier-Stokes equations are
\begin{align}
    \partial_{t}\mathbf{u}+\mathbf{u}\cdot\nabla\mathbf{u}+\nabla p&=\nu\nabla^{2}\mathbf{u}-a\mathbf{u}+\mathbf{f},\label{BEQ1}\\
    \nabla\cdot\mathbf{u}&=0,\label{BEQ2}
\end{align}
where the components of velocity $\mathbf{u}$ are bounded by $u_{m}$, notion $p$ stands for the pressure, function $\mathbf{f}$ denotes the external forcing, and $a$ denotes the drag coefficient \cite{boffetta2012two}. For freely decaying turbulence, there is no forcing effect, i.e., $\mathbf{f}\equiv \mathbf{0}$. For forced turbulence, we define $\mathbf{f}$ as the classic Kolmogorov forcing with $4$ forcing waves \cite{rollin2011variations}. In all simulations, we set $a=0.1$. 

By defining a scalar vorticity field 
\begin{align}
\omega=\nabla\times\mathbf{u}\label{BEQ3}\end{align}
and a stream function $\psi$ such that 
\begin{align}
\mathbf{u}=\left(\partial_{x}\psi,-\partial_{y}\psi\right)\label{BEQ4},
\end{align}
we can reformulate Eqs. (\ref{BEQ1}-\ref{BEQ4}) as \cite{boffetta2012two}
\begin{align}
\partial_{t}\omega+\mathbf{u}\cdot\nabla\omega=\nu\nabla^{2}\omega-a\omega+\nabla\times\mathbf{f}\label{BEQ5}.
\end{align}
We set a periodic boundary condition for Eq. (\ref{BEQ5}) and solve it applying the Fourier pseudo-spectral method \cite{hussaini1987spectral,sengupta2021analysis}. The time advancement is realized by a Crank-Nicolson Runge-Kutta method of order $4$ and the time resolution is automatically optimized following the Courant–Friedrichs–Lewy condition to ensure numerical stability \cite{Kochkov2021-ML-CFD,Dresdner2022-Spectral-ML}. See Fig. \ref{AG0} for instances.

After generating vorticity trajectories under each condition of $\left(\nu,\mu_{m}\right)$, we sample $50$ equally spaced time steps and save the corresponding vorticity fields into data. Meanwhile, we can combine Eq. (\ref{BEQ4}) and the Fourier transform of 
\begin{align}
\nabla^{2}\psi=-\omega\label{BEQ6}
\end{align}
to map a vorticity field to its associated velocity field if necessary (see instances in Fig. \ref{AG0}).

\section{Generation of fluid networks}\label{ASec4}
In our work, we define fluid networks based on vorticity fields to keep consistency with Refs. \cite{nair2015network,taira2016network} (i.e., vortical networks). Given a scalar field $\omega$ at specific moment (i.e., a $256\times 256$ matrix), we first down-sample it into a spatial resolution of $64\times 64$. The down-sampling is implemented in a naive manner, where we subdivide the original $256\times 256$ matrix into $64\times 64$ patches and average vorticity values within each patch to derive a mean vorticity value representing this patch. The obtained $64\times 64$ scalar field is used to define a complete graph of $64^{2}$ units, where each pair of units share an edge weighted according to the induced
velocity (i.e., the iterations among fluid elements) \cite{nair2015network,taira2016network}. Specifically, we apply a symmetric weight assignment scheme, i.e., the un-directed edge between units $i$ and $j$ is weighted as 
\begin{align}
W_{ij}&=\frac{1}{2}\left(\frac{\vert\omega_{ij}\vert}{2\pi\vert x_{i}-x_{j}\vert}+\frac{\vert\omega_{ji}\vert}{2\pi\vert x_{j}-x_{i}\vert}\right)\label{CEQ1},
\end{align}
where $\omega_{ij}$ denotes the $\left(i,j\right)$-th element of matrix $\omega$ and each $x_{i}\in D$ denotes the spatial coordinate of patch $i$. Then, we filter out those edges whose weights are blow $\kappa$, a percentile of $0.8$ (i.e., smaller than $80\%$ of weights). The filtering leads to the final adjacency matrix of fluid network
\begin{align}
A_{ij}&=\left(1-\delta_{k}\left(i,j\right)\right)\Theta\left(W_{ij}-\kappa\right)W_{ij}\label{CEQ2},
\end{align}
where $\delta_{k}\left(\cdot,\cdot\right)$ stands for the Kronecker delta function and $\Theta\left(\cdot\right)$ denotes the unit step function.

There are $3\times 6\times 50=900$ networks corresponding to each kind of fluid flow, which are used for subsequent analysis. One can see examples in Fig. \ref{AG0}.

\section{Measurement of fluid properties}\label{ASec5}
In our work, we use kinetic energy and enstrophy to characterize fluid properties \cite{kundu2015fluid}.

\paragraph*{Kinetic energy.} Given a velocity field $\mathbf{u}$, we can measure its kinetic energy as
\begin{align}
E=\frac{1}{2}\int_{D}\mathbf{u}_{x}\cdot \mathbf{u}_{x}\mathsf{d} x.\label{DEQ1}
\end{align}
\paragraph*{Enstrophy.} Meanwhile, we can measure the enstrophy of the corresponding vorticity field $\omega$
\begin{align}
\Omega=\frac{1}{2}\int_{D}\omega_{x}\omega_{x}\mathsf{d} x.\label{DEQ2}
\end{align}

Apart from kinetic energy and enstrophy, there are several auxiliary variables used in our fluid analysis, whose definitions are presented below.

\paragraph*{Integral length scale.} To characterize the largest possible motions of a fluid flow, we can calculate the intergral length scale \cite{boffetta2012two,kundu2015fluid}
\begin{align}
L=\frac{u^{*}}{\omega^{*}},\label{DEQ3}
\end{align}
where $u^{*}$ denotes the root mean square velocity (i.e., a single scalar)
\begin{align}
u^{*}=\sqrt{\frac{1}{4\pi^{2}}\int_{D}\mathbf{u}_{x}\cdot \mathbf{u}_{x}\mathsf{d} x}\label{DEQ4}
\end{align}
and $\mathbf{u}^{*}$ is the root mean square vorticity
\begin{align}
\omega^{*}=\sqrt{\frac{1}{4\pi^{2}}\int_{D}\omega_{x}\omega_{x}\mathsf{d} x}.\label{DEQ5}
\end{align}
In Eqs. (\ref{DEQ4}-\ref{DEQ5}), the area of domain $D$ is measured as $4\pi^{2}$.

\paragraph*{Reynolds number.}To reflect the dynamic ratio between dissipation and fluid gradient growth, we can consider the Reynolds number
\begin{align}
Re=\frac{u^{*}L}{\nu},\label{DEQ6}
\end{align}
where $\nu$ is the viscosity defined before.

Note that all these fluid properties are measured in the original velocity and vorticity fields rather than their down-sampled counterparts. 

\section{Measurement of fluid network properties}\label{ASec6}
The macroscopic observables used for characterizing fluid networks are selected as weighted degree distribution \cite{taira2016network}, degree-degree distance distribution \cite{meng2023scale,zhou2020power}, and network control cost \cite{klickstein2017energy,klickstein2018control}.

\paragraph*{Weighted degree distribution.} The weighted degree distribution is defined as \cite{taira2016network}
\begin{align}
P\left(\operatorname{deg}=n\right)=\frac{1}{N}\sum_{i}\delta\left(A_{ij}-n\right),\label{EEQ1}
\end{align}
where $\operatorname{deg}$ denotes the weighted degree, notion $N$ measures the number of units, and $\delta\left(\cdot\right)$ is the Dirac delta function.

\begin{figure*}[!t]
\includegraphics[width=1\columnwidth]{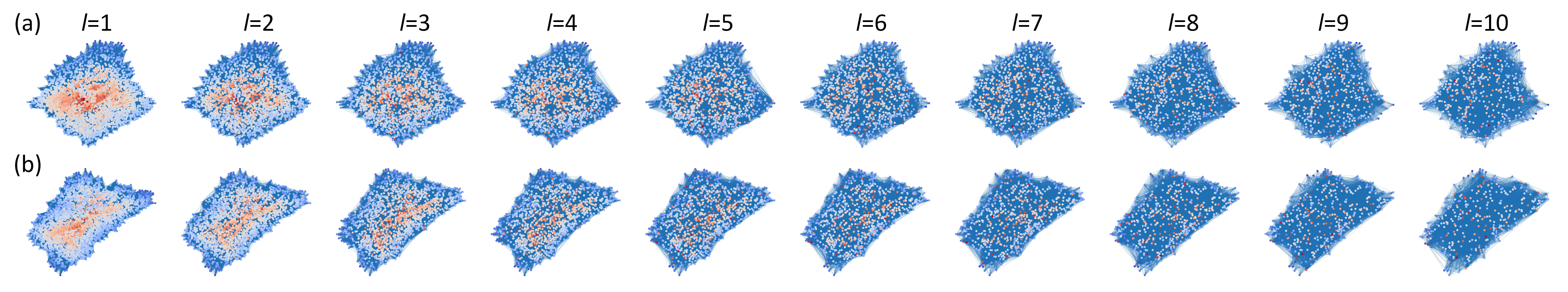}
\caption{\label{AG1} Two instances of the renormalization flows of fluid networks. (a-b) respectively show the renormalization flows of the decaying (a) and forced (b) turbulence defined in Fig. \ref{AG0}. Initial fluid networks are generated at moment $t=0$.} 
\end{figure*}

\paragraph*{Degree-degree distance distribution.} Similarly, the degree-degree distance distribution is defined as \cite{meng2023scale,zhou2020power}
\begin{align}
P\left(\eta=n\right)&=\frac{1}{M}\sum_{i}\delta\left(\eta-n\right),\label{EEQ2}\\
\eta\left(i,j\right)&=\exp\left(\Bigg\vert\log\frac{\operatorname{deg}\left(i\right)}{\operatorname{deg}\left(j\right)}\Bigg\vert\right),\;\forall \left(i,j\right)\in \text{Edges},\label{EEQ3}
\end{align}
where $\eta\left(i,j\right)$ reflects the connectivity imbalance between the two endpoints of an edge $\left(i,j\right)$ and $M$ measures the number of edges.

\paragraph*{Control cost.} To measure the cost of controlling a fluid network, we follow Refs. \cite{klickstein2017energy,klickstein2018control} to consider the linear controller problem widespread in engineering practices. Specifically, we analyze the following linear invariant dynamic system
\begin{align}
\partial_{t}\mathbf{s}\left(t\right)&=A\mathbf{s}\left(t\right)+B\mathbf{i}\left(t\right),\label{EEQ4}\\
\mathbf{o}\left(t\right)&=C\mathbf{s}\left(t\right),\label{EEQ5}
\end{align}
where $\mathbf{s}\left(t\right)\in\mathbb{R}^{N}$ denotes the state vector of $N$ units affected by a control input vector $\mathbf{i}\left(t\right)\in\mathbb{R}^{N}$. Vector $\mathbf{o}\left(t\right)\in\mathbb{R}^{N}$ is the output vector, i.e., control target. Notion $A\in \left(0,\infty\right)^{N\times N}$ denotes the adjacency matrix defined in Eq. (\ref{CEQ2}), notion $B\in \{0,1\}^{N\times m}$ is the injection matrix of input signals, and $C\in \{0,1\}^{p\times N}$ is the output matrix of control targets. Same as Ref. \cite{klickstein2017energy}, we consider a case where each column of $B$ and $C$ contains only a single non-zero element (i.e., each signal component is processed by one input unit and transmitted to one target unit). If a system described by Eqs. (\ref{EEQ4}-\ref{EEQ5}) is controllable during an interval of $\left[0,t^{\prime}\right]$, there exists an optimal input $\widehat{\mathbf{i}}\left(t\right)$ that can drive
the system from an initial condition, $\mathbf{s}\left(0\right)$, to a desired output, $\mathbf{o}\left(t^{\prime}\right)$, with a minimum cost (see Refs. \cite{klickstein2017energy,klickstein2018control} for the expression of $\widehat{\mathbf{i}}\left(t\right)$ in details). The minimum cost is measured as \cite{klickstein2017energy,klickstein2018control}
\begin{align}
E_{c}=J^{\top}\left(CF\left(t^{\prime}\right)C^{\top}\right)^{-1}J,\label{EEQ6}
\end{align}
where $J$ quantifies the control action 
\begin{align}
J=\mathbf{o}\left(t^{\prime}\right)-C\exp\left(At^{\prime}\right)\mathbf{s}\left(0\right)\label{EEQ7}
\end{align}
and $F\left(t^{\prime}\right)$ denotes the controllability Gramian
\begin{align}
F\left(t^{\prime}\right)=\int_{0}^{t^{\prime}}\exp\left(A\left(t^{\prime}-\tau\right)\right)BB^{\top}\exp\left(A^{\top}\left(t^{\prime}-\tau\right)\right)\mathsf{d}\tau.\label{EEQ8}
\end{align}
The control action defined by Eq. (\ref{EEQ7}) reflects the effective difference between $\mathbf{o}\left(t^{\prime}\right)$ and the zero-input output at moment $t^{\prime}$. Eq. (\ref{EEQ6}) quantifies the minimum cost of realizing such a control action \cite{klickstein2017energy,klickstein2018control}. In the case with single input and single control target, i.e., $m=p=1$, matrices $B$ and $C$ reduce to two vectors with single non-zero elements, respectively. The corresponding minimum cost has a simple upper bound when the infinite path network approximation is applied (see details in Ref. \cite{klickstein2018control})
\begin{align}
E_{c}\simeq J_{\ell}^{2}\left[\int_{0}^{\infty}\exp\left(-2q\tau\right)I_{\ell}^{2}\left(2f\tau\right)\mathsf{d}\tau\right]^{-1},\label{EEQ9}
\end{align}
where $\ell$ denotes the location of the non-zero element in vector $C$, notion $J_{\ell}$ stands for the $\ell$-th element of $J$, parameter $f$ assigns the weights of edges during control, and $q$ quantifies self-regulation effects \cite{klickstein2017energy}. Function $I_{\ell}\left(\cdot\right)$ is the modified Bessel function of the first
Kind of integer order
\begin{align}
I_{\ell}\left(z\right)=\frac{1}{2\pi}\int_{0}^{\pi}\exp\left(z\cos\left(\theta\right)\right)\cos\left(\ell\theta\right)\mathsf{d}\theta.\label{EEQ10}
\end{align}
In our work, Eq. (\ref{EEQ9}) is implemented to measure the control costs of fluid networks. Without loss of generality, we assume that the non-zero element of $B$ is always located at the $1$-st component. Thus, index $\ell$ also represents the distance between control target (i.e., the unit who generates system output) and input (i.e., the unit who receives signal input) under the infinite path network approximation \cite{klickstein2018control}. Meanwhile, we set $f=1$ for convenience and define $q=\lambda_{\text{max}}+1$ following Ref. \cite{klickstein2018control}, where $\lambda_{\text{max}}$ denotes the largest eigenvalue of fluid network adjacency matrix, $A$ (here $q=\lambda_{\text{max}}+1$ actually makes the largest eigenvalue of matrix $A-qE$ equal $-1$ and prevents the control action from being diverge, in which $E$ denotes an identity matrix \cite{klickstein2018control}). 

\begin{figure*}[!t]
\includegraphics[width=1\columnwidth]{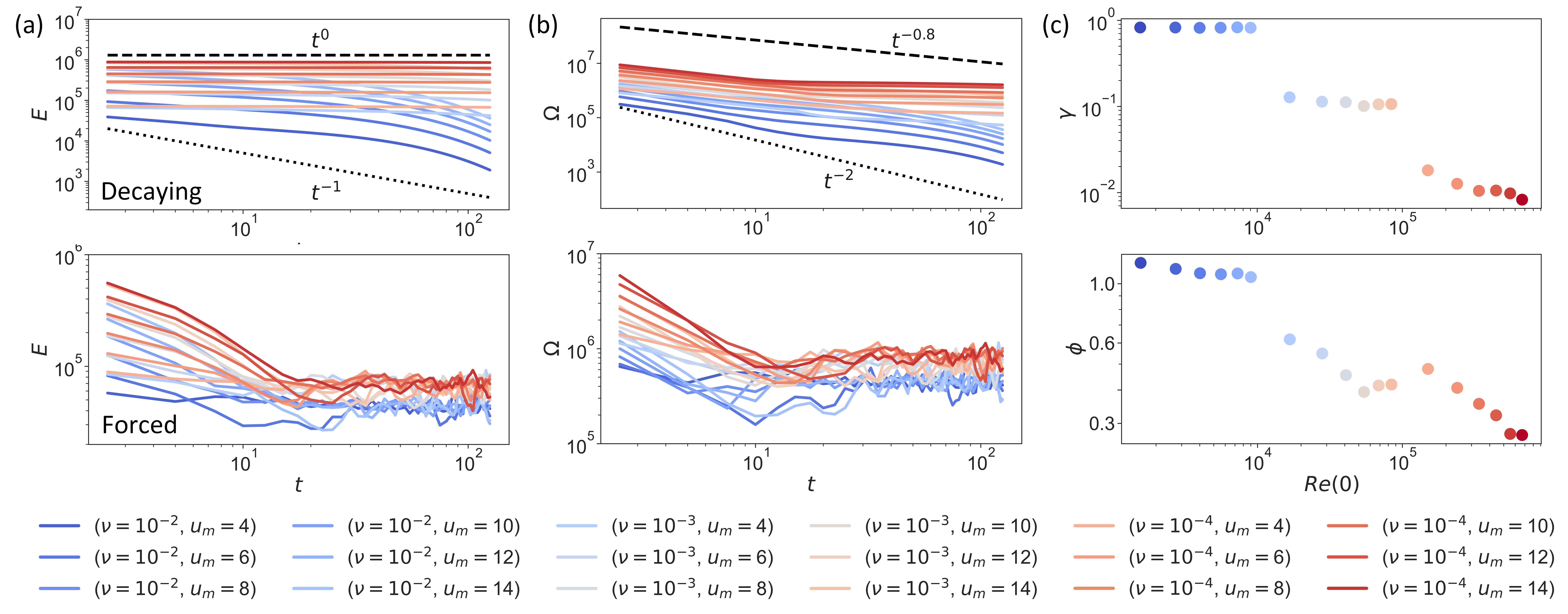}
\caption{\label{AG2} Basic results of fluid analysis. (a) The decay trends of kinetic energy of freely decaying and forced turbulent flows are shown. The dashed line denotes the case with an extremely large Reynolds number, where energy remains essentially constant \cite{chasnov1997decay}. The doted line denotes the normal case where energy decreases following $t^{-1}$ \cite{chasnov1997decay}. (b) The decay trends of enstrophy of decaying and forced turbulence are illustrated. The dashed line corresponds to an extremely large Reynolds number case, where enstrophy follows $t^{-0.8}$ \cite{chasnov1997decay}. The doted line denotes the normal case where enstrophy usually behaves as $t^{-2}$ \cite{chasnov1997decay}. (c) The decay law exponents, $\gamma$ and $\phi$, of kinetic energy and enstrophy in decaying turbulence are shown as the functions of the initial Reynolds number, $Re\left(0\right)$ (i.e., the Reynolds number at moment $0$), respectively.} 
\end{figure*}

\section{Renormalization group}\label{ASec7}
To analyze the scaling behaviours of fluid networks, we apply the random renormalization group (RRG) proposed in our earlier work \cite{tian2024fast}. The RRG is a general framework established on random projections, hashing techniques, and kernel representations, which supports fast unfolding of ultra-large systems (e.g., with millions of units) within minutes. Different from classic model-based renormalization flows used in turbulence analysis \cite{yakhot1986renormalization,nagano1997renormalization,mizerski2020renormalization}, the RRG enables us to renormalize the structures (e.g., networks) and dynamics of complex systems in a unified and model-free manner. Therefore, the RRG naturally fits in with our demands of processing a large set of fluid networks.

The theoretical and technical details of the RRG can be seen in Ref. \cite{tian2024fast} and the code implementation is offered in Ref. \cite{RRGrelease}. Below, we sketch the key steps and parameter settings of the RRG pipeline for fluid network renormalization.

Given a fluid network, $X$, consisting of $N$ units, we set it as the input of the RRG, i.e., $X^{\left(1\right)}=X$. In each $l$-th iteration ($l\geq 1$), we progressively realize the following procedures:
\begin{itemize}
\item[(1) ] We define a feature representation, $Y^{\left(l\right)}$, of $X^{\left(l\right)}$ such that every unit $X^{\left(l\right)}_{i}$ corresponds to a feature vector $Y^{\left(l\right)}_{i}$. Specifically, we let $W^{\left(l\right)}_{i}$ be the set of unit $X^{\left(l\right)}_{i}$ and all its adjacent units. Then, we apply the MinHash method \cite{broder1997resemblance,broder1998min} to hash $W^{\left(l\right)}_{i}$ as $Y^{\left(l\right)}_{i}$ such that $Y^{\left(l\right)}_{i}$ captures the local adjacency properties of unit $X^{\left(l\right)}_{i}$ in a space of desired dimension.
\item[(2) ] We use the signed Cauchy projection \cite{li2013sign} to hash $Y^{\left(l\right)}$ into a binary form $Z^{\left(l\right)}$ such that the correlation distance between $Y^{\left(l\right)}_{i}$ and $Y^{\left(l\right)}_{j}$ in the Cauchy kernel space can be approximated by the Hamming distance between $Z^{\left(l\right)}_{i}$ and $Z^{\left(l\right)}_{j}$. In the analysis, each $Z^{\left(l\right)}_{i}$ is set to have a dimension of $20$.
\item[(3) ] We implement an approximate nearest neighbor search on $Z^{\left(l\right)}$ to identify the nearest neighbor of every unit $X^{\left(l\right)}_{i}$. These nearest neighbor relations are included in space $U^{\left(l\right)}$.
\item[(4) ] We generate a null network, $G^{\left(l\right)}$, of all units and gradually add edges into it. Specifically, an edge is added between each pair of nearest neighbors in $U^{\left(l\right)}$ only if they are adjacent in network $X^{\left(l\right)}$ as well. After adding all possible edges, every connected cluster, $C^{\left(l\right)}_{k}$, is made up by the units with strong correlations.
    \item[(5) ] Finally, we renormalize all the units in each connected cluster $C^{\left(l\right)}_{k}$ into a macro-unit $X^{\left(l+1\right)}_{k}$. Two macro-units are defined with an edge in $X^{\left(l+1\right)}$ if the units aggregated into them have at least one edge in $X^{\left(l\right)}$.
    \end{itemize}

    By iterating these five steps, the RRG progressively generates a renormalization flow, $\left(X^{\left(1\right)},\ldots,X^{\left(T\right)}\right)$, of the fluid network, where $T$ denotes the iteration number. In our experiment, we set $T=15$. See Fig. \ref{AG1} for the instances of fluid network renormalization.
    
    One thing to note is how we define edge weights during renormalization. If two macro-units, $X^{\left(l\right)}_{i}$ and $X^{\left(l\right)}_{j}$, share an edge, then the weight of this edge is averaged across all the edge weights leading from the initial units recursively aggregated into $X^{\left(l\right)}_{i}$ to the initial units recursively aggregated into $X^{\left(l\right)}_{j}$. Here initial units refer to the original units contained in $X^{\left(1\right)}$ \cite{RRGrelease}.

\section{Analysis of fluid decay laws}\label{ASec8}
In this work, we consider a power-law form decay of the kinetic energy across time
\begin{align}
    E\propto t^{-\gamma}\label{GEQ1},
\end{align}
where $\gamma$ approaches to $0$ when the Reynolds number is extremely large \cite{chasnov1997decay}. In most normal cases, we expect to see $\gamma=1$ \cite{chasnov1997decay}.

    \begin{figure*}[!t]
\includegraphics[width=1\columnwidth]{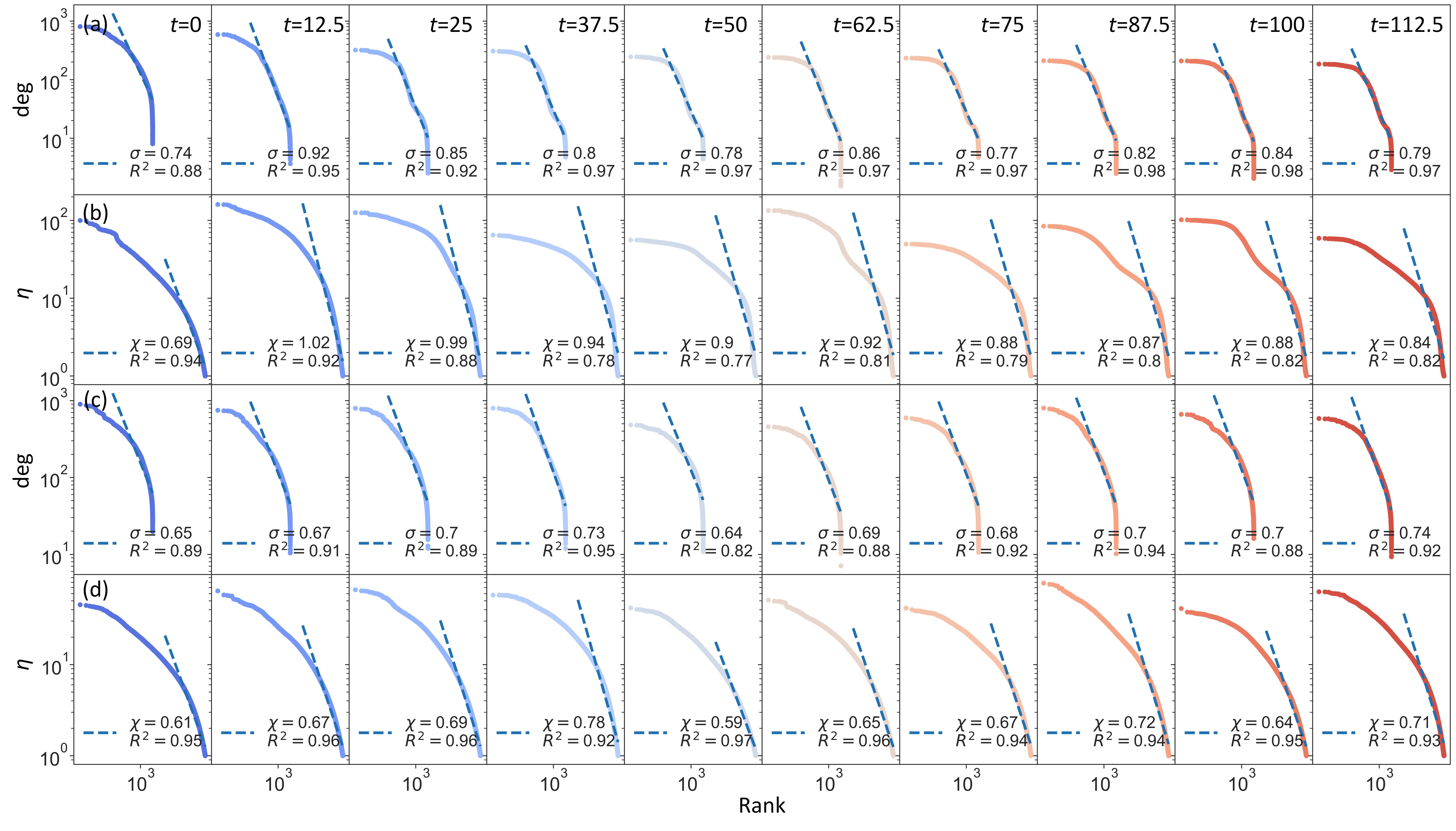}
\caption{\label{AG3} Two instances of spatial self-similarity analysis. (a-b) The estimations of power-law exponents, $\sigma$ and $\chi$, are shown on the decaying turbulence data generated in Figs. \ref{AG0}(a-d). (c-d) The same analysis is implemented on the forced turbulence data generated in Figs. \ref{AG0}(e-h). Note that estimation accuracy is quantified by the R square value, $R^{2}$. Scatters denote empirical observations and dashed lines denote fitted models.} 
\end{figure*}

Meanwhile, we consider a similar decay of the enstrophy 
\begin{align}
    \Omega\propto t^{-\phi}\label{GEQ2},
\end{align}
where $\phi=0.8$ frequently occurs given an extremely large Reynolds number \cite{chasnov1997decay}. In other normal cases, we usually observe $\phi=2$ \cite{chasnov1997decay}.

Note that the exponents of these decay laws are estimated directly using the least square method in a log-log space rather than the approach introduced in Sec. \ref{ASec7} because the latter only applies to power-law probability distributions \cite{clauset2009power,virkar2014power}. The derived results in our experiments are presented in Fig. \ref{AG2}.

\begin{figure*}[!t]
\includegraphics[width=1\columnwidth]{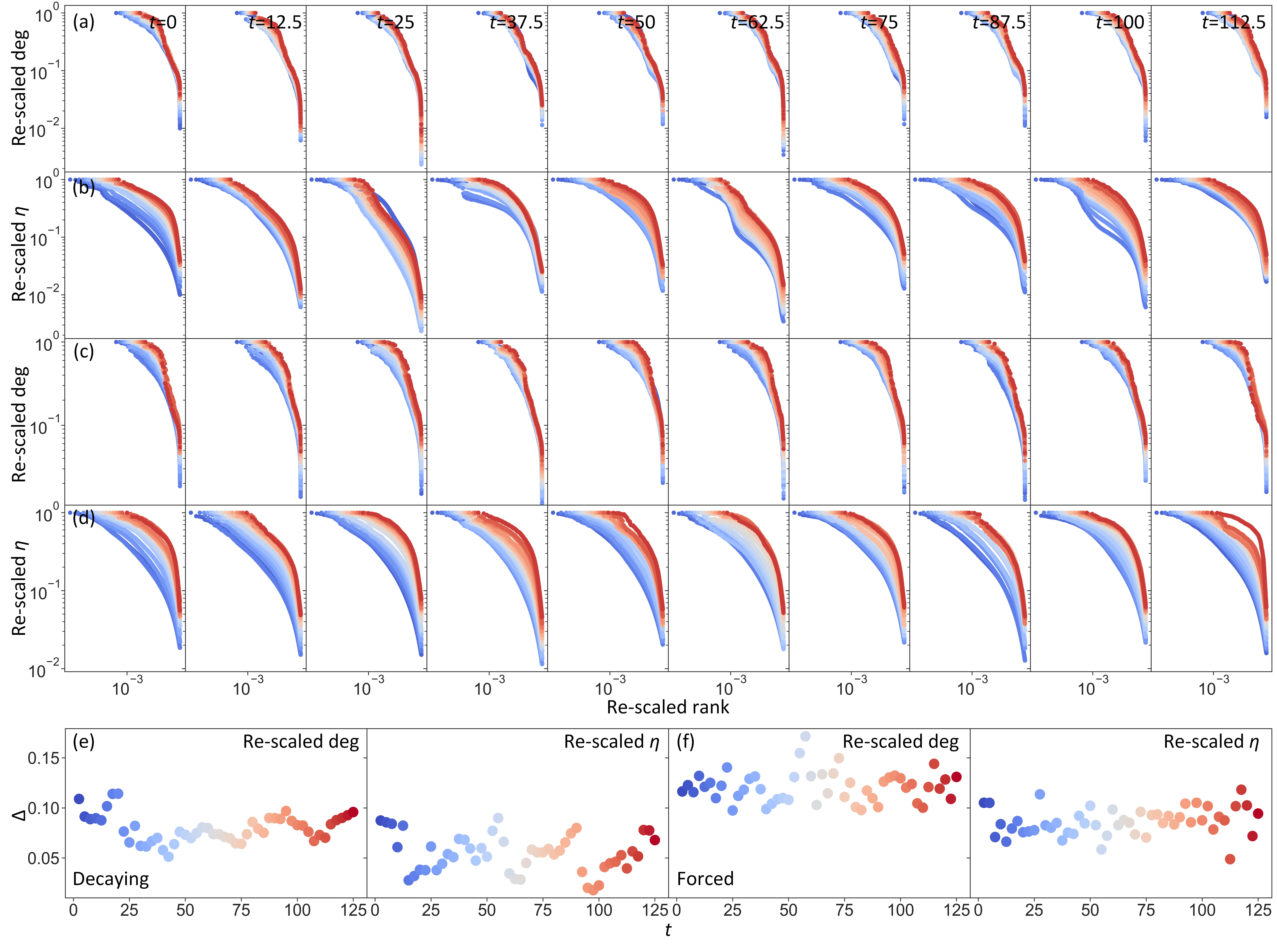}
\caption{\label{AG4} Two instances of scale-invariance analysis. (a-b) The renormalization flows of the decaying turbulence data generated in Figs. \ref{AG0}(a-d) are shown, where the selected macroscopic observables are weighted degree and degree-degree distance, respectively. The color of scatters, changing from red to blue, represents the enlarging iteration number of renormalization. (c-d) The same analysis is implemented on the forced turbulence data in Figs. \ref{AG0}(e-h). (e-f). The measured evolution intensities of the renormalization flows shown in (a-d) change across time.} 
\end{figure*}

\section{Analysis of spatial self-similarity and its generalization}\label{ASec9}
To verify the existences of classic and generalized spatial self-similarity, we need to statistically test if probability distributions $P\left(\operatorname{deg}=n\right)$ and $P\left(\eta=n\right)$ in Eqs. (\ref{EEQ1}-\ref{EEQ3}) exhibit power-law behaviours. In previous studies \cite{clauset2009power,virkar2014power}, this task has been solved by a combination of maximum likelihood estimation and semi-parametric bootstrap test. However, the high computational and sample complexities of this approach essentially limit its applicability to our ultra-large and non-smooth data sets (e.g., there are $1679579$ sparsely distributed samples of $\eta$ in each single case). As an alternative solution, here we suggest a new way to verify these power-law probability distributions. Below, we summarize its key ideas by taking $P\left(\eta=n\right)$ as an instance.

Our method is rooted a simple fact about the power-law probability distribution. If $P\left(\eta=n\right)$ exhibits power-law behaviours
\begin{align}
P\left(\eta=n\right)=V\eta^{-e},\label{HEQ1}
\end{align}
where $V$ stands for a proper normalization term and $e$ is the power-law exponent, we can reformulate Eq. (\ref{HEQ1}) as
\begin{align}
\frac{1}{M}\int_{0}^{\infty}\delta\left(\eta-g\left(r\right)\right)\mathsf{d}r\simeq V\eta^{-e}\label{HEQ2}
\end{align}
by assuming the continuity. In Eq. (\ref{HEQ2}), parameter $M$ measures the sample number of $\eta$ as mentioned in Eq. (\ref{EEQ2}). Function $g\left(\cdot\right)$ is a mapping from a rank number to $\eta$, i.e., sample $\eta_{r}=g\left(r\right)$ is the $r$-th largest value among all samples. In practice, we can directly sort the sample set to derive $g\left(\cdot\right)$ numerically.

\begin{figure*}[!t]
\includegraphics[width=1\columnwidth]{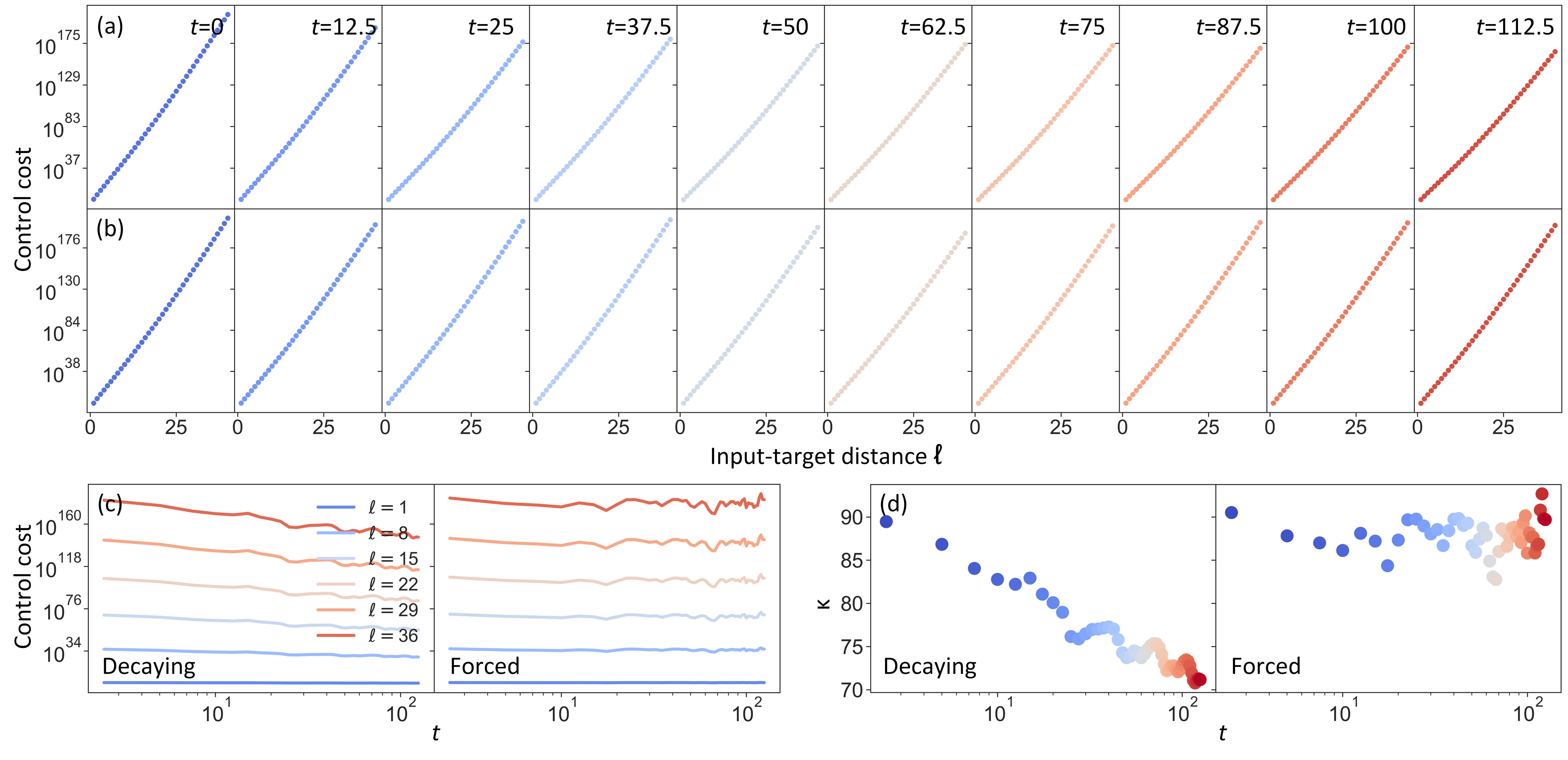}
\caption{\label{AG5} Two instances of control scaling analysis. (a-b) The control costs of the decaying (a) and forced turbulent flows (b) generated in Fig. \ref{AG0} exhibit exponential scaling behaviours with respect to $\ell$, the distance between input and control target. (c) The control costs associated with different $\ell$ are shown as the functions of time. (d) The control scaling exponent, $\kappa$, is shown as a function of time.} 
\end{figure*}

For convenience, we define $h\left(\eta,r\right)=\eta-g\left(r\right)$ and $r_{\eta}=g^{-1}\left(\eta\right)$. Because the Dirac delta function satisfies
\begin{align}
\delta\left(h\left(\eta,r\right)\right)=\frac{\delta\left(r-r_{\eta}\right)}{\vert \partial_{r} h\left(\eta,r_{\eta}\right)\vert},\label{HEQ3}
\end{align}
we can transform Eq. (\ref{HEQ2}) into
\begin{align}
\frac{1}{M}\frac{1}{\vert \partial_{r} h\left(\eta,r_{\eta}\right)\vert}\int_{1}^{\infty}\delta\left(r-r_{\eta}\right)\mathsf{d}r&\simeq V\eta^{-e},\label{HEQ4}\\
\frac{1}{M}\frac{1}{\vert \partial_{r} h\left(\eta,r_{\eta}\right)\vert}&\simeq V\eta^{-e}.\label{HEQ5}
\end{align}
Meanwhile, we know $\partial_{r} h\left(\eta,r\right)=-\partial_{r} g\left(r\right)$ and $h\left(\eta,r\right)$ is a monotone increasing function whose zero point is $r_{\eta}$. Therefore, Eq. (\ref{HEQ5}) is equivalent to 
\begin{align}
-\frac{1}{M}\frac{1}{\partial_{r} g\left(r_{\eta}\right)}&\simeq Vg\left(r_{\eta}\right)^{-e}.\label{HEQ6}
\end{align}
Solving Eq. (\ref{HEQ6}), we can know
\begin{align}
g\left(r\right)&=\left(xr+y\right)^{\frac{1}{1-e}},\label{HEQ7}\\
\eta_{r}&\propto r^{\frac{1}{1-e}},\label{HEQ8}
\end{align}
where $x$ and $y$ are specific constants. As shown in Eq. (\ref{HEQ8}), the power-law behaviours of probability distribution $P\left(\eta=n\right)$ create a power-law rank scaling phenomenon of $\eta$, i.e., $\eta_{r}\propto r^{-\chi}$ with $\chi=\frac{1}{e-1}$. In fact, it is easy follow a similar idea to prove that the power-law rank scaling of $\eta$ also implies a power-law probability distribution $P\left(\eta=n\right)$.

Consequently, we can analyze the rank scaling of samples to indirectly verify the power-law probability distribution of the concerned random variable. In our work, the power-law exponent $\chi$ of degree-degree distances is estimated using a least square method, where the fitting region is determined as $r\in\left[0.01M,0.99M\right]$ (i.e., covering $98\%$ samples) to control the effects of extreme values. Meanwhile, the same approach is also applied to degrees to derive their power-law exponent $\theta$, i.e., $\operatorname{deg}_{r}\propto r^{-\sigma}$. One can see Fig. \ref{AG3} for instances of estimated results.

\section{Analysis of scale-invariance}\label{ASec10}
 To verify the scale-invariance property, we first choose a macroscopic observable as the analysis target. For fluid networks, either weighted degree \cite{taira2016network} or degree-degree distance \cite{meng2023scale,zhou2020power} can be selected. A fluid network is scale-invariant if the chosen macroscopic observable, after being re-scaled, remains invariant during renormalization (i.e., the system is born at the fix point of renormalization flow). Here re-scaling is implemented to control the order of magnitude of the macroscopic observable via linear transformations. Taking degree-degree distance as an instance, we suggest to consider a re-scaling mapping
 \begin{align}
\left(r,\eta_{r}\right)\mapsto \left(\bar{r},\bar{\eta}_{\bar{r}}\right):=\left(\frac{r}{r_{m}},\frac{\eta_{r}}{\max_{r}\eta_{r}}\right)\in\left[0,1\right]^{2},\label{IEQ1}
\end{align}
where $r_{m}$ denotes the maximum rank. It is easy to verify that the original power-law rank scaling still holds
 \begin{align}
\bar{\eta}_{\bar{r}}\propto \bar{r}^{-\chi}.\label{IEQ2}
\end{align}
Therefore, the re-scaling mapping in Eq. (\ref{IEQ1}) does not change data trends. For convenience, we denote $\left(\bar{r},\bar{\eta}_{\bar{r}}\right)$ as the re-scaled macroscopic observable in subsequent derivations.

After excluding the effects of data magnitudes, we can measure the evolution intensity of the macroscopic observable during renormalization as
\begin{align}
\Delta=\big\langle \Delta\left(l,1\right)\big\rangle_{l},\label{IEQ3}
\end{align}
where $\langle\cdot\rangle_{l}$ denotes the averaging across $l\in\{1,\ldots,T\}$ and $\Delta\left(l,1\right)$ measures the departure of the macroscopic observable in the $l$-th iteration from the one in the $1$-th iteration
\begin{align}
    \Delta\left(l,1\right)=\Big\langle \Big\vert\bar{\eta}_{\bar{r}}^{\left(l\right)}-\bar{\eta}_{\bar{r}}^{\left(1\right)}\Big\vert\Big\rangle_{\bar{r}}.\label{IEQ4}
\end{align}
In Eq. (\ref{IEQ4}), we refer to $\left(\bar{r}^{\left(l\right)},\bar{\eta}_{\bar{r}}^{\left(l\right)}\right)$ as the re-scaled macroscopic observable derived using $X^{\left(l\right)}$, the fluid network in the $l$-th iteration of the renormalization flow. Eq. (\ref{IEQ4}) calculate the average absolute difference between $\bar{\eta}_{\bar{r}}^{\left(l\right)}$ and $\bar{\eta}_{\bar{r}}^{\left(1\right)}$ across all possible values of $\bar{r}$ to quantify the departure intensity. Given a scale-invariant system, the value of $\Delta$ in Eq. (\ref{IEQ3}) is expected to approach to $0$. Otherwise, the value of $\Delta$ increases to $1$. The above approach can be directly applied on weighted degrees \cite{taira2016network} as well. Please see Fig. \ref{AG4} for examples.
 
Certainly, one can also consider other kinds of scale-invariance tests (e.g., a two-sided Kolmogorov–Smirnov test \cite{simard2011computing,berger2014kolmogorov} between the distributions of macroscopic observable across different iterations of renormalization \cite{tian2024fast}). Eqs. (\ref{IEQ1}-\ref{IEQ4}) only serve as practical ways to evaluate scale-invariance property on large-scale data sets.

\section{Analysis of control scaling}\label{ASec11}
As shown in Ref. \cite{klickstein2018control}, the minimum control cost under the infinite path network approximation scales following an exponential function of $\ell$
\begin{align}
E_{c}\propto \exp\left(\kappa \ell\right),\label{JEQ1}
\end{align}
where $\ell$, as we mentioned before, represents the distance between signal input and control target \cite{klickstein2018control} and $\kappa$ is the scaling exponent estimated using the least square method.

In Fig. \ref{AG5}, we present two instances of control scaling analysis. As shown in these results, the control costs of decaying turbulence associated with different $\ell$ gradually decrease across time while the control costs of forced turbulence generally maintain robust.

 \bibliography{apssamp}
	\newpage


\end{document}